\documentclass{SCIS2026}
\usepackage{fontawesome5}
\usepackage{xcolor}

\hypersetup{
    colorlinks=true,
    linkcolor=blue,
    citecolor=blue
}
\usepackage{orcidlink}
\usepackage[capitalise, nameinlink]{cleveref}
\begin{document}
\Year{2026}
\Month{January}
\Vol{68}
\No{1}
\DOI{}
\ArtNo{}
\ReceiveDate{}
\ReviseDate{}
\AcceptDate{}
\OnlineDate{}       
\AuthorMark{}
\AuthorCitation{}

\title{A Token/KV-Cache Communication Media Selection and Resource Allocation Strategy for Multi-Agent Collaboration}
{Dai L P, Xiang L P, Yang K. A token/KV-cache communication media selection and resource allocation strategy for multi-agent collaboration}

\author[1,2]{Lipeng Dai\textsuperscript{\scalebox{1.2}{\orcidlink{0000-0002-1924-262X}}\,}}{}
\author[1,2]{Luping Xiang\textsuperscript{\scalebox{1.2}{\orcidlink{0000-0003-1465-6708}}\,}}{}
\author[1,2]{Kun Yang}{}

\address[1]{State Key Laboratory of Novel Software Technology, Nanjing University, Nanjing, 210008, China}
\address[2]{Institute of Intelligent Networks and Communications (NINE), Nanjing University (Suzhou Campus), Suzhou, 215163, China}
\address[]{Emails: dlp1022@163.com, luping.xiang@nju.edu.cn, kunyang@nju.edu.cn}

\abstract{
    The convergence of large language models (LLMs) with 6G networks is fostering a paradigm of autonomous multi-agent cooperation, which in turn is expected to substantially increase east-west traffic. 
    Although latent-space interaction mechanisms can enable more efficient collaboration than symbolic natural-language (NL) exchanges, prior work often abstracts away the associated communication overhead under practical wireless constraints. In embodied multi-agent settings, heterogeneous interaction media incur disparate inference and transmission costs, thereby inducing an inherent end-to-end (E2E) latency trade-off. To address this, we propose a joint design that integrates communication-media selection with wireless resource allocation. Through analytical characterization and simulation-based evaluation, we show that neither token-based transmission nor key-value (KV) cache-based transmission is uniformly optimal across operating regimes, as performance depends critically on system parameters such as available computational resources and channel conditions. Accordingly, we formulate a joint optimization problem aimed at minimizing the E2E latency of multi-agent collaboration and develop a low-complexity joint media selection and resource allocation (JMSRA) algorithm. Numerical results further confirm that, by adaptively coordinating the interaction media and bandwidth allocation over heterogeneous links, the proposed scheme achieves markedly reduced E2E latency relative to conventional NL-only and KV-cache-only baselines, enabling efficient and robust multi-agent collaboration in future wireless networks.
}

\keywords{Multi-agent collaboration, large language models, wireless networks, communication media, bandwidth allocation}

\maketitle

\section{Introduction}

For several decades, the design of mobile communication networks has been largely oriented toward human-centric services, yielding a predominantly north-south traffic profile in which data are exchanged mainly between cloud platforms and end devices. 
With the emergence of the 6G vision, however, AI workloads are progressively migrated from centralized clouds to the network edge to satisfy stringent latency and reliability requirements~\cite{EdgeAI6G2022_JSAC}, {and enable pervasive AI-as-a-service~\cite{OverviewAI6G2025_SCI}}.
This shift redistributes traffic toward localized computing nodes and mitigates congestion in the core network. 
Moreover, the rapid proliferation of embodied agents~\cite{EmbodiedMAS2025_JAS} and agentic AI~\cite{AgenticAI6G2025_CM} in applications such as autonomous driving and industrial automation is accelerating a more profound transition—from base-station-mediated communication to decentralized end-to-end (E2E) interactions. 
It is further projected that the global population of active AI agents will reach the trillion scale by 2036, exceeding the number of conventional connected devices by over two orders of magnitude~\cite{AgentNumber2025_arXiv}.

Within this evolving landscape, a key vision is the realization of autonomous and seamless inter-agent interaction. 
Such collaboration may fundamentally reshape human-machine ecosystems by reducing the reliance on manual supervision and enabling embodied robotic systems to cooperate in executing complex tasks. 
Achieving this objective presupposes reliable inter-agent communication. 
In multi-embodied-agent systems, the exchange of internal states, sensory observations, and action intentions over wireless links is essential for effective collaboration. 
In particular, to support heterogeneous agents with diverse capabilities and representations, the choice of an appropriate interaction approach is critical—analogous to human communication via speech, text, or gesture.

{
In the context of large language model (LLM)-driven multi-agent systems (MASs), the interaction approach is defined by the choice of \textit{communication media}. 
Here, the concept of a communication medium transcends the conventional physical wireless link, elevating to the semantic and cognitive level. 
Specifically, when an LLM-driven agent intends to convey its intentions to others, it can choose to transmit either generated text or internal neural features. 
Based on this fundamental distinction, communication media in current multi-agent collaboration are primarily categorized into two paradigms: the natural language (NL) paradigm and the latent space (LS) paradigm.}

At present, NL remains the dominant interaction medium for LLM-driven MAS~\cite{EntireLatentComm2025_arXiv,ThoughtComm2025_arXiv,CacheToCache2025_arXiv,LatentCollab2025_arXiv,QKVComm2025_arXiv,AgentPrimitives2026_arXiv}, which typically rely on the exchange of discrete tokens. 
Nevertheless, discrete token exchanges may be suboptimal for certain collaboration tasks, as compact nonverbal signals can sometimes convey richer intent than verbose textual descriptions. 
From this perspective, the reliance on NL can bound the representational expressiveness of agents, echoing Wittgenstein’s observation that the limits of language delimit the limits of one’s world~\cite{wittgenstein2023tractatus}. 
Similar to human communication, which is susceptible to ambiguity and misinterpretation, NL-based agent collaboration may be impaired by linguistic uncertainty, while deeper latent cognitive content may remain inexpressible within symbolic token sequences.

Fundamentally, NL is a construct optimized for human social exchange. 
For intelligent agents, this is not a necessary constraint, and they may instead benefit from communication directly in their internal representation space. 
Motivated by this premise, recent efforts have investigated leveraging the LS of LLMs, e.g., key-value (KV) cache representations and latent ``thoughts", as alternative interaction media for MASs~\cite{EntireLatentComm2025_arXiv,ThoughtComm2025_arXiv,CacheToCache2025_arXiv,LatentCollab2025_arXiv,QKVComm2025_arXiv,AgentPrimitives2026_arXiv}. 
Reported results indicate that LS-based interaction can improve collaboration efficiency and task success rates relative to NL-based messaging. In particular, cache-to-cache communication can reduce computational overhead by enabling downstream agents to reuse intermediate computations produced by upstream agents.

Despite these benefits, LS-based interaction introduces nontrivial costs. 
Unlike NL, which enjoys high-level symbolic compression, LS mechanisms require the exchange of high-dimensional latent variables, particularly for KV-cache transmission~\cite{QKVComm2025_arXiv}. 
Existing studies in the natural language processing community often assume agents are co-located within high-performance computing clusters and therefore understate or neglect the corresponding communication overhead. 
In contrast, embodied agents in physical environments must interact over stochastic wireless channels with stringent resource constraints. 
In such settings, transmitting large tensors can incur prohibitive latency, potentially offsetting the computational advantages of LS-based collaboration and rendering theoretical gains impractical. 
A summary comparison between NL and LS paradigms is provided in~\cref{tab:paradigm_comparison}.

\begin{table}[!t]
    \footnotesize
    \caption{Comparison of LLM-Agent Communication Paradigms}
    \label{tab:paradigm_comparison}
    \tabcolsep 0pt 

    \begin{tabular*}{\textwidth}{@{\hspace{2cm}} l @{\extracolsep{\fill}} l l @{\hspace{2cm}}}
        \toprule
        \textbf{Metric} & \textbf{NL} & \textbf{LS} \\
        \midrule
        Data Representation & Discrete Token IDs & Continuous Tensors \\
        Bandwidth Demand    & Low (Variable with length) & Extremely High (Raw FP16) \\
        Information Density & Low (Syntactic Redundancy) & High (Concentrated) \\
        Computational Cost  & High (Full Enc/Decoding) & Low (Direct Injection) \\
        Generality          & High (Cross-architecture) & Low (Model-specific) \\
        Noise Robustness    & Moderate (Self-correcting) & Low (Sensitive to Noise) \\
        \bottomrule
    \end{tabular*}
\end{table}

{To follow the energy and latency requirements of sustainable 6G~\cite{AISustainable6G2025_SCI}, the communication overhead of LLM-based multi-agent collaboration must be balanced against the limited resource budgets of the edge devices.}
Inspired by the aforementioned trade-offs, in the complex and stochastic landscape of 6G wireless communications, embodied agents are faced with a delicate tug-of-war between computational savings and communication overhead. 
Therefore, a fundamental, yet unexplored question arises:

\begin{center}
    {\color{orange}\faLightbulb} \emph{\textbf{Can we dynamically navigate the trade-off between computation and communication latency to determine the optimal medium for multi-agent collaboration in practical environments?}}
\end{center}

To address this question, this paper proposes a joint mode selection and resource allocation (JMSRA) framework for 6G-enabled embodied MASs. 
{Unlike conventional delay minimization studies in edge networks (e.g., \cite{DelayLoad2025_TNSM}), which primarily focus on task queuing and node trajectory planning, the JMSRA tackles the latency bottleneck by treating the communication medium itself as a strategic variable.} 
Rather than adhering to a static communication protocol, our framework empowers agents to adaptively switch between NL and LS paradigms by perceiving real-time wireless channel fluctuations and computational constraints. 
Besides, to further enhance system performance, the framework optimizes wireless resource allocation based on the selected communication mode and current network conditions.
Our key contributions are summarized as follows:
\begin{itemize}
    \item We conducted a mathematical analysis of E2E latency (including computation and transmission) in a hierarchical MAS with two-round conversation.
          Through simulations, we revealed the dynamic shift in advantages between the token-based and KV-cache-based transmission modes, demonstrating that neither is universally optimal as system parameters change.
    \item To minimize E2E latency, we derived the closed-form communication media selection strategy from the perspective of a single sending agent. 
          This allows the agent to autonomously decide whether to transmit tokens or KV cache based on the real-time assessment of the current system state (including itself, the wireless channel, and the receiving agent).
    \item To investigate communication media selection in multi-agent wireless transmission scenarios featuring bandwidth resource competition, we formulated a mixed-integer non-linear programming (MINLP) problem to jointly optimize bandwidth allocation and media selection vectors.
          To solve this, we proposed a low-complexity JMSRA algorithm.
          Numerical results demonstrate that the proposed scheme optimally orchestrates the selection of communication media and allocates the most appropriate bandwidth for agents based on system parameters, including LLM and wireless channel characteristics, leading to significantly lower E2E latency compared to baselines and maintaining excellent stability in dynamic multi-round dialogue scenarios.
\end{itemize}

\section{Related Work}

\subsection{Multi-Agent Collaboration in Latent Space}
To transcend limitations of NL in multi-agent collaboration, recent studies have focused on the LS paradigm. 
For instance, \cite{EntireLatentComm2025_arXiv} introduced Interlat, a mind-reading framework that leverages the last hidden states of LLMs for direct transmission, enabling agents to externalize their nuance reasoning beyond textual constraints. 
\cite{ThoughtComm2025_arXiv} introduced ThoughtComm.
By treating agent communications as identifiable latent variables, this work offers formal proofs that agents can achieve mutual understanding through neural states alone, bypassing the need for traditional symbolic supervision.
Furthermore, several works focus on leveraging the KV cache as a medium for semantic transfer. 
\cite{CacheToCache2025_arXiv} enabled direct semantic communication by mapping the source model's KV-cache into the target model's internal layers via a learnable alignment mechanism, effectively bypassing the redundant token-by-token re-encoding latency.
\cite{LatentCollab2025_arXiv} proposed an E2E latent collaboration mechanism that utilizes a shared latent working memory to ensure lossless information exchange, achieving higher expressiveness with lower complexity than vanilla text-based MAS. 
To address the efficiency of such high-dimensional exchange, \cite{QKVComm2025_arXiv} introduced an adaptive quantization and hybrid information extraction protocol, achieving significant compression ratios while maintaining semantic fidelity.
{To mitigate error accumulation and latency in long-context interactions, \cite{AgentPrimitives2026_arXiv} abstracted MAS architectures into reusable primitives that communicate internally entirely via KV cache, significantly reducing token consumption.}

\subsection{LLM in Wireless Communication}
As LLMs migrate from centralized clouds to the network edge, researchers have begun exploring wireless communication strategies for LLMs.
{Against the backdrop of LLM development, \cite{TowardsWAIM2025_SCIS} proactively explored the key elements, technical characteristics, and development methodology of wireless big AI model.}
To promote the development of LLM in edge networks, \cite{EdgeInferenceLLM2025_TWC} proposed a edge LLM inference framework and maximized the throughput in a wireless system with multi-user and multi-edge nodes. 
To minimize total service latency, \cite{EfficientLLMInference2025_arXiv} proposed an LLM serving framework based on speculative decoding, where small models rapidly generate draft tokens while large models perform parallel verification. 
Besides, for distributed deployment, the communication overhead of frequent tensor synchronization remains a critical concern. 
To address this, \cite{OverTheAirLLM2025_arXiv} utilized over-the-air computation to leverage the analog superposition property of wireless multiple-access channels, effectively mitigating the communication bottleneck of tensor parallelism in on-device LLM inference.
To mitigate backhaul pressure, \cite{CachingInferenceLLM2025_ICC} jointly orchestrated model caching and resource scheduling, effectively exploring the trade-off between local edge processing and remote cloud inference.
Complementary to the transmission of inference results, AirNet proposed in~\cite{AirNet2024_TWC} pioneered the direct mapping of neural network parameters to channel symbols by employing structured pruning and unequal error protection.
{
    Futhermore, when extending LLMs to physical-layer forecasting, \cite{FASLLM2026_JSAC} proposed a novel architecture that embeds compressed delay-Doppler representations into a LoRA-adapted LLM, proving its capability to achieve low-latency sequence modeling on wireless channels.
}

\section{System Model}
\label{sec:system_model}

\begin{figure}[!h]
    \centering
    \includegraphics[width=0.68\textwidth]{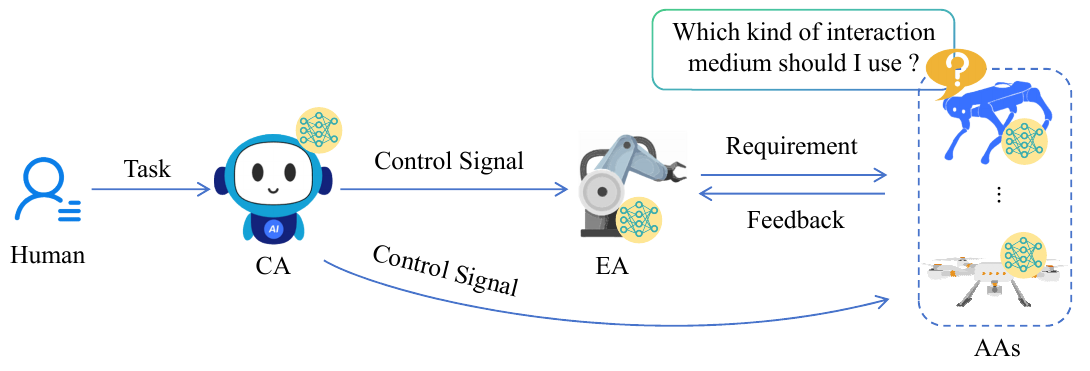}
    \caption{Embodied agents collaboration system.}
    \label{fig:system_model}
\end{figure}

\subsection{Hierarchical Cooperation Architecture}
\label{sec:hierarchical_architecture}
The architecture of MASs has two primary types are considered: sequential and hierarchical~\cite{LatentCollab2025_arXiv,BeyondSelfTalk2025_arXiv}. 
Sequential MASs rely on hop-by-hop relaying, causing the total latency to scale linearly with the number of agents and increasing E2E uncertainty. 
Given the stringent latency requirements of wireless collaborative agent systems, we adopt the hierarchical architecture.
As illustrated in~\cref{fig:system_model}, upon receiving a task from human, a central agent (CA) assigns the task to an executive agent (EA) and orchestrates several assistant agents (AAs) to help it. 
The EA assumes the role of a local summarizer to lead the collaboration and perform final physical actions, while the AAs, characterized by heterogeneous knowledge, serve as specialized experts, e.g., an unmanned aerial vehicle (UAV) providing multi-view inference.
{
To orchestrate the collaborative interaction within the MAS, the CA acts as a centralized coordinator with a global perspective, maintaining real-time knowledge of the global channel quality, the internal states of each agent, and other relevant parameters to formulate and dispatch control directives to all agents.
}

The collaborative signal flow is structured as follows. 
Initially, the EA performs preliminary semantic processing of the task prompt, multicasts its initial output to all AAs via wireless channels, and then wait for their responses. 
Upon receiving the data, each AA decodes the EA's intent and invokes its local multimodal sensors, such as cameras or radar, to acquire real-time environmental data. 
Subsequently, the AAs conduct further reasoning by integrating the newly acquired environmental insights with the inherited context, and transmit the updated findings back to the EA via orthogonal frequency-division multiple access (OFDMA).
These feedback messages encapsulate insights derived from each AA's unique sensory perspective or specialized knowledge base. 
For instance, in a logistics scenario where a robotic arm is assigned to grab a package, it then relies on one UAV to provide critical environmental data, such as the relative spatial coordinates and potential obstacles, which are beyond the robotic arm's local sensing range. 
Finally, the robotic arm performs a summarizing inference by synthesizing these multi-perspective insights with its internal state to generate the final control commands for grasping and placing the package.

\subsection{Decoder-only Transformer Architecture}
\label{sec:decoder_only_transformer}
We assume that each agent is modeled as a entity comprising a pre-trained LLM, perception/tool modules, and a single-antenna.
Then, this section provides a mathematical description of the Transformer architecture employed in these agents.
The evolution of Transformer-based models has yielded three predominant architectural paradigms: the encoder-decoder, the encoder-only , and the decoder-only architecture. 
This paper assumes the agents are equipped with the decoder-only architecture, which serves as the backbone for classic LLMs such as the GPT, Qwen, and LLaMA series.
One of the core of this architecture is the causal masked self-attention mechanism, which ensures that the generation of a target token is strictly conditioned on the preceding sequence~\cite{LLMsurvey2023_arXiv}.
Stemming from this unidirectional autoregressive characteristic, the inference process leverages the KV cache technique to substantially enhance computational efficiency.

\begin{figure}[!h]
    \centering
    \includegraphics[width=0.75\textwidth]{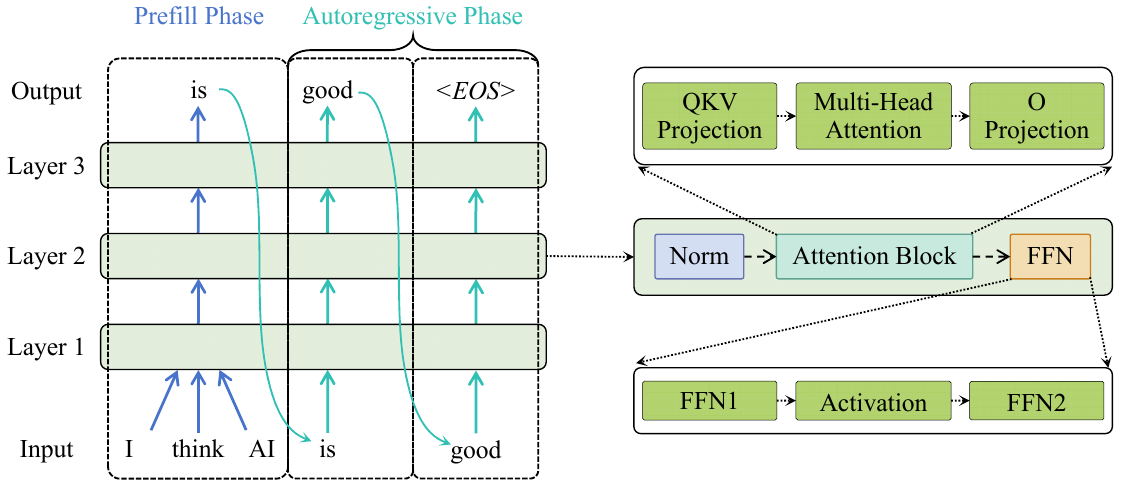}
    \caption{Inference procedure of LLM with a decoder-only Transformer~\cite{OrcaOSDI22}.}
    \label{fig:Inference_procedure}
\end{figure}

A typical decoder-only Transformer layer and its corresponding inference procedure are illustrated in~\cref{fig:Inference_procedure}.
Specifically, the signal flow is bifurcated into two stages: the prefill phase and the autoregressive decoding phase.

\emph{1) Prefill Phase:} During the initial phase, the entire input sequence is processed in parallel, allowing each token to attend to all input tokens through the self-attention mechanism.
Mathematically
\footnote{
For the sake of simplicity, the additive bias terms and details like LayerNorm are omitted in this paper. 
This is justified by the fact that the computational workload in high-dimensional Transformer layers is predominantly governed by matrix multiplications~\cite{EdgeInferenceLLM2025_TWC}.
}
, the agent processes an input prompt with $s$ tokens and transforms it into an embedding matrix. 
Let $\mathbf{X}_l \in \mathbb{R}^{s \times d_m}$ denote the corresponding input matrix of the $l$-th Transformer block, where $d_m$ represents the hidden dimensionality. 
For each of the total $L$ Transformer blocks, its input latent representation is projected into query ($\mathbf{Q}_l$), key ($\mathbf{K}_l$), and value ($\mathbf{V}_l$) matrices to compute attention matrix $\mathbf{A}_l = \text{softmax}\left(\frac{\mathbf{Q}_l \mathbf{K}_l^T}{\sqrt{d_h}}\right)\mathbf{V}_l$, where $\mathbf{Q}_l,\, \mathbf{K}_l,\, \mathbf{V}_l = \mathbf{X}_l \mathbf{W}_{Q,\,l},\, \mathbf{X}_l\mathbf{W}_{K,\,l},\, \mathbf{X}_l\mathbf{W}_{V,\,l}$.
Moreover, $\mathbf{W}_{Q,\,l},\, \mathbf{W}_{K,\,l},\, \mathbf{W}_{V,\,l} \in \mathbb{R}^{d_m \times (Hd_h)}$ are learnable weight matrices, where $H$ and $d_h$ are the number of attention heads and the dimensionality of each head, respectively.
Next, $\mathbf{A}_l \mathbf{W}_{O,\,l}$ is transmitted to the feed-forward network (FFN) for further processing, and then we obtain the $l$-th layer's output $\mathbf{Y}_l = \sigma(\mathbf{A}_l \mathbf{W}_{O,\,l} \mathbf{W}_{f1,\,l}) \mathbf{W}_{f2,\,l}$, where $\mathbf{W}_{O,\,l} \in \mathbb{R}^{(Hd_h) \times d_m}$ is the output projection matrix, $\mathbf{W}_{f1,\,l} \in \mathbb{R}^{d_m \times d_{f}}$ and $\mathbf{W}_{f2,\,l} \in \mathbb{R}^{d_{f} \times d_m}$ the learnable FFN parameters, $d_f$ the FFN expansion factor, and $\sigma(\cdot)$ the activation function.
At last, the last row vector $\mathbf{y}_s \in \mathbb{R}^{1 \times d_m}$ of $\mathbf{Y}_L$ is projected onto the vocabulary space through $logits = \mathbf{y}_s \mathbf{W}_p$ for the prediction of the first output token, where $\mathbf{W}_p \in \mathbb{R}^{d_m \times d_v}$ is the vocabulary projection matrix and $d_v$ the vocabulary size.

\emph{2) Autoregressive Phase:} In this phase, the model generates the remaining $\alpha-1$ tokens, with each new token generation conditioned on all previously generated tokens.
With the aforementioned KV cache mechanism, the signal flow at this stage exhibits a typical incremental computation pattern.
To generate the $n$-th token ($n \in \left\{2, 3, \cdots, \alpha\right\}$), the input vector $\mathbf{x}_{n-1} \in \mathbb{R}^{1 \times d_m}$ is derived from the embedding of the previously generated token.
Subsequent operations mirror the prefill phase, with the critical distinction that only the query, key, and value vectors $\mathbf{q}, \mathbf{k}, \mathbf{v} \in \mathbb{R}^{1 \times (H d_h)}$ for the current step are computed newly. 
The newly generated KV pair is then appended to the historical cache, represented as $\mathbf{K} = [\mathbf{K}_{cache}; \mathbf{k}]$, $\mathbf{V} = [\mathbf{V}_{cache}; \mathbf{v}]$.

{It is worth noting that this autoregressive inference is not merely a computation procedure, but the very foundation for communication media selection. 
Technically, as will be further analyzed in \cref{sec:latent_space_mode}, in a pure LS interaction paradigm, the model can bypass the final vocabulary projection step and directly feed the last layer's hidden states as the input for the next autoregressive step, meaning the generation of tokens is not strictly mandatory. 
To enable agents to dynamically switch communication media in fluctuating wireless environments, we stipulate that tokens must be generated during the inference process.
This ensures that both representational modalities are simultaneously available in the local memory, thereby rendering them as parallel options at the moment of transmission. 
This duality establishes the premise for the dynamic media selection in the subsequent sections.}

\section{Latency Analysis}
\label{sec:latency_analysis}
From the perspective of communication systems, inter-agent coordination is fundamentally a process of the encoding, transmission, and decoding of information.
Given that signal design directly governs the system's performance, communication strategies must be tailored to the specific characteristics of the underlying information source. 
In the context of LLM-based MASs, this tailoring involves a strategic choice between the NL and LS paradigms.
To evaluate their effectiveness, this section {first establishes two static baselines.}
{Specifically, we formulate the E2E latency for the two-round dialogue between the EA and AAs under two extreme cases: an all-NL mode (where all agents exclusively transmit tokens) and an all-KV mode (where all agents exclusively transmit KV caches).
To ensure a fair comparison, we keep the task workloads (e.g., LLM input and output lengths) strictly identical in both modes.}
Subsequently, their performance is compared through simulations.
{Results demonstrate that neither mode is universally superior, which naturally motivates the dynamic media selection proposed in \cref{sec:optimal_mode_selection}.}

\subsection{Token Transmission Mode}
\label{sec:token_mode}
In token-based communication mode, the E2E latency is primarily composed of two parts: the computational latency for inference at the EA and AAs, and wireless transmission latency.

Following the rules in~\cref{fig:Inference_procedure}, we assume that each agent receives an input prompt of $s_{i,j}$ tokens and generates $\alpha_{i,j}$ output tokens\footnote[2]{
    Here, $\alpha_{i,j}$ counts the output tokens excluding the final end-of-sequence terminator $\langle EOS \rangle$.
}
, where $i \in \{0, 1, \cdots, I \}$, $i = 0$ denotes the EA, $i > 0$ denotes the $I$ AAs, and $j \in \{ 1, 2 \}$ denotes the conversation round\footnote[3]{
    Consistent with \cref{sec:hierarchical_architecture}, each participating agent executes two distinct inference rounds.
}
.
Particularly, the input token number of EA at the initial stage is denoted by $s_{0,1} = s$, where $s$ is the initial prompt size.
At the final stage, the number changes to $s_{0,2} = \sum_{i=1}^{I} \alpha_{i,2}$.


Focusing on {latency}, {the actual processing delay in practical LLM-based systems is primarily dictated by memory I/O (e.g., reading model weight and KV cache) and inference computation.
For cloud or edge servers, computational resources are abundant, and thus they are typically I/O-bound.
To address severe I/O bottlenecks, state-of-the-art parallel scheduling techniques (e.g., FlexGen~\cite{FlexGenICML23}, Cake~\cite{pmlr-v267-jin25d}, and DualPath~\cite{dualpath2026_arXiv}) are proposed, thereby effectively reducing the I/O latency so that it approaches, and can even fall below, the inference latency.
Conversely, for embodied agents, they are typically equipped with limited computational resources, and thus they are often computation-bound.
Therefore, we consider decoupling the end-to-end delay from I/O fluctuations and use the inference latency of standard decoder-only architectures as the metric for system performance evaluation\footnote[1]{
    {While recent algorithmic advancements in state-of-the-art models, such as the mixture-of-experts and sparse attention mechanisms adopted by DeepSeek~\cite{DeepSeek2025_arXiv}, reduce theoretical floating-point operations, the basic inference latency in standard decoder-only architectures can be used as an upper-bound baseline.}
}.}

Specifically, inference latencies can be approximated by the time required for multiplying large matrices during the prefill and autoregressive phases, which are respectively given by\footnote[2]{
  The computational workload is quantified in terms of floating-point operations (FLOPs). For the product of two matrices $\mathbf{A} \in \mathbb{R}^{a \times c}$ and $\mathbf{B} \in \mathbb{R}^{c \times b}$, the total workload is estimated as $2abc$ FLOPs, accounting for both multiplication and addition operations.
}
\begin{align}
    \label{eq:inference_latency_natural_language1}
    &t_{\text{NL,pr},i,j} (s_{i,j}, \varphi_{i,j}) = \frac{1}{C_i} \left\{ L \left[ 6  H d_h d_m s_{i,j} + 4 H d_h \varphi_{i,j} s_{i,j}   + (2 H d_h d_m s_{i,j} + 4 d_m d_f s_{i,j}) \right] + 2 d_m d_v \right\}, \\
    \label{eq:inference_latency_natural_language2}
    &t_{\text{NL,au},i,j} (\alpha_{i,j}, \varphi_{i,j}) = \frac{\alpha_{i,j}}{C_i} \left\{ L \left[ 6 H d_h d_m + 4 H d_h (\varphi_{i,j} + \frac{1 + \alpha_{i,j}}{2}) + ( 2 H d_h d_m + 4 d_m d_f ) \right] + 2 d_m d_v \right\},
\end{align}
where $C_i$ represents the computational capacity\footnote[3]{
    {To accurately characterize the computing capacity of embodied agents, we define $C_i = \eta_i C_{i,\text{peak}}$, where $C_{i,\text{peak}}$ denotes the theoretical peak computational capacity dictated by the specific type of hardware accelerator employed (e.g., GPU or NPU), and $\eta_i \in (0, 1]$ represents the currently available computing capacity coefficient governed by the agent's real-time power constraints.}
}
, 
{measured in floating-point operations per second (FLOPS)},
$\varphi_{i,j} = s_{i,j} + \theta_{i,j}$, and $\theta_{i,j}$ is the historical context length accumulated prior to the input of these $s_{i,j}$ tokens.
In \cref{eq:inference_latency_natural_language1,eq:inference_latency_natural_language2}, the three terms from left to right within the square brackets represent the computational costs of QKV projections, the attention matrix calculation, and the FFN output operations, respectively.
Owing to the linear growth of KV cache during autoregressive decoding, the term $(1 + \alpha_{i,j})/2$ serves as a metric for the average computational complexity.
Furthermore, the rightmost term in the curly braces denotes the complexity of projecting latent representations onto the vocabulary space.
Consequently, the total inference latency of each agent is given by
\begin{align}
    t_{\text{NL,inf},i,j} (\alpha_{i,j}, \varphi_{i,j}, s_{i,j})  = \frac{k_1 \alpha_{i,j}^2 + 2 k_1 \varphi_{i,j} (\alpha_{i,j} + s_{i,j}) + ( k_1 + k_2 + k_3 ) \alpha_{i,j} + k_2 s_{i,j} + k_3}{C_i},
\label{eq:inference_latency_natural_language_total}
\end{align}
where $k_1$ to $k_3$ are constants determined by the model parameters in each agent, $k_1 = 2 L H d_h$, $k_2 = 8 L H d_h d_m + 4 L d_m d_f$, $k_3 = 2 d_m d_v$.


Focusing on the wireless transmission latency, it is primarily influenced by the transmitted token number, the characteristics of the wireless channel, and other related factors.
Departing from traditional cellular communication method, this paper considers a direct agent-to-agent (A2A) communication approach, which can be accomplished through 6G sidelink technology~\cite{6GSidelink2021_CSM}.
Let $b$ denote the number of bits representing each token index, typically determined by $b = \lceil \log_2 d_v \rceil$.
Furthermore, let $h_{\text{AE},i}$ and $h_{\text{EA},i}$ represent the channel gains for the link between the $i$-th AA and the EA in both directions, respectively. 
Given the transmit power $P_i$ and the total bandwidth $B$, the communication rate that the EA broadcasts to all AAs and that from the $i$-th AA to the EA are formulated by
\begin{align}
    \label{eq:R0}
    R_0 &= \min_{i \in \{1, \cdots, I\}} B \log_2(1 + \frac{P_0 h_{\text{EA},i}^2}{B N_0}), \\
    \label{eq:Ri}
    R_i &= \rho_i B \log_2(1 + \frac{P_i h_{\text{AE},i}^2}{\rho_i B N_0}), \quad i \in \{1, \cdots, I\}, 
\end{align}
where $N_0$ is the power spectral density of the additive white Gaussian noise (AWGN), and $\rho_i$ is the bandwidth allocation coefficient for the $i$-th AA.
Then, the corresponding transmission latencies are formulated as
\begin{align}
    \label{eq:communication_latency_natural_language_e}
   t_{\text{NL,com},0}  &= \frac{b \alpha_{0,1}}{R_0}, \\
    \label{eq:communication_latency_natural_language_w}
   t_{\text{NL,com},i}  &= \frac{b \alpha_{i,2}}{R_i}, \quad i \in \{1, \cdots, I\}.
\end{align}

In summary, the total E2E latency of the token-based NL paradigm can be expressed as
\begin{align}
    t_{\text{NL}} = \sum_{j = 1}^{2} t_{\text{NL,inf},0,j} +t_{\text{NL,com},0} + \max_{i \in \{1,2, \cdots, I \} } \left\{ \sum_{j=1}^{2} t_{\text{NL,inf},i,j} +t_{\text{NL,com},i} \right\}.
    \label{eq:total_latency_natural_language}
\end{align}

\subsection{KV Transmission Mode}
\label{sec:latent_space_mode}

\begin{figure}[!h]
    \centering
    \includegraphics[width=0.5\textwidth]{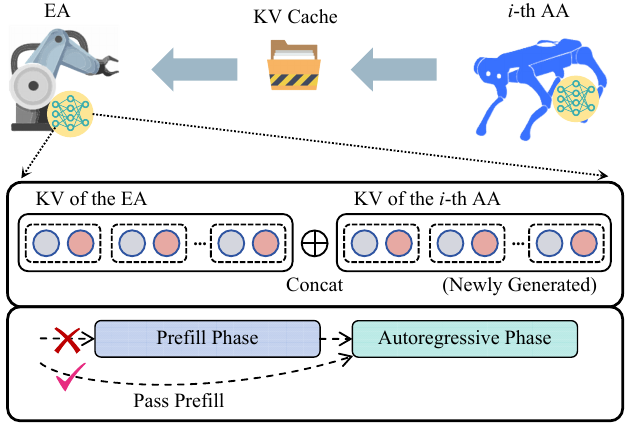}
    \caption{KV communication between two LLM-agents.}
    \label{fig:KV_communication}
\end{figure}

In scenarios where multi-agent collaboration occurs without human intervention, the interaction media need not be restricted to human-readable language. 
As illustrated in~\cref{fig:KV_communication}, transmitting KV cache, which encapsulates the model's rich semantic understanding, enables agents to collaborate more effectively by leveraging their deep cognitive representations. 
Transmitting KV cache from agent A to agent B can reduce redundant computations in agent B by inheriting the contextual memory already established by agent A.
This leads to the fact that agent B can bypass the prefill phase and directly enter the autoregressive decoding stage.
However, this approach necessitates the transmission of high-dimensional tensors over wireless channels, which may introduce significant communication overhead.
Therefore, we next analyze the E2E latency of this mode from both computational and communication perspectives.

Before delving into the analysis of E2E latency between agents, we first need to determine the internal reasoning mechanisms of the agents. 
As studied in~\cite{LatentCollab2025_arXiv}, the hidden representation of the last layer of the Transformer model can be directly used as input for subsequent steps in the autoregressive phase.
Thus, the token prediction steps can be passed.
This latent-variable driven inference ensures that the underlying semantic nuances are not lost during the autoregressive process.
For agents, they can choose either discrete tokens or latent variables as the input during the autoregressive phase.
If discrete tokens are chosen, agents must bear the computational overhead incurred by the token-prediction layer. 
Conversely, if latent variables are chosen as the input, an input-output alignment\footnote[1]{
    This is due to the input layer of LLM is trained for token embeddings.
}
operation is required to ensure semantic consistency, which can be achieved through $\mathbf{x}_t = \mathbf{y}_{t-1} \mathbf{W}_{\text{align}}$, where $\mathbf{W}_{\text{align}} = (\mathbf{W}_{\text{out}}^T \mathbf{W}_{\text{out}} + \lambda \mathbf{I})^{-1} \mathbf{W}_{\text{out}}^T \mathbf{W}_{\text{in}}$ is an alignment projector, $\mathbf{W}_{\text{out}}$ and $\mathbf{W}_{\text{in}}$ the output and input embedding layers~\cite{LatentCollab2025_arXiv}.
It can be seen that both methods incur additional computational overhead during the autoregressive phase. 
In fact, the computational cost of token prediction is marginally lower than that required for latent distribution alignment.
To ensure a fair comparison with the token-based mode, we opt to generate tokens as outputs and utilize them as inputs for subsequent steps. 
Furthermore, this approach facilitates the inspection of system logs by human operators, enabling them to identify and rectify potential system failures more effectively.

Next, driven by the computation-communication trade-off of the KV cache collaboration mode, we provide a quantitative analysis of the E2E latency for this mode in the subsequent content.

The E2E latency under this mode is determined by two complete and two partial inference, alongside two wireless transmission delays. 
In particular, the EA's initial inference latency for the task prompt is equivalent to its counterpart in the NL mode at the same stage, such that 
\begin{align}
    \label{eq:EA_KV_inference_latency1}
    t_{\text{KV,inf},0,1} = t_{\text{NL,inf},0,1}.
\end{align}

Next, we determine the data volume of the broadcasted KV cache from the EA.
Since the KV cache consists of floating-point numbers, we assume the use of FP16 storage precision. 
Given that FP16 utilizes 16 bits to encode each value, a matrix $A \in \mathbb{R}^{a \times b}$ requires $16ab$ bits for transmission.
While the standard FP16 representation serves as a baseline, the resulting wireless transmission overhead remains substantial. 
To mitigate this, advanced quantization techniques can be employed to compress the KV cache~\cite{QKVComm2025_arXiv,QAQ_2025_ICCV}. 
For instance, as suggested in~\cite{QKVComm2025_arXiv}, layer-wise mixed-precision quantization, which allocates 8 bits to sensitive layers and 4 bits to less sensitive ones, can significantly reduce data volume at the cost of slight accuracy degradation. 
For analytical tractability, we assume a uniform quantization scheme in this paper and introduce a compression ratio $\gamma_i$. 
For example, $\gamma_0 = 2$ indicates a reduction from the original 16 bits to 8 bits per element at the EA.
Then, the bit number transmitted by the EA can be expressed as
\begin{align}
    \label{eq:KV_cache_size1}
    m_{\text{KV},0} = 32 L H d_h \frac{ s_{0,1} + \alpha_{0,1} }{\gamma_0} .
\end{align}
Therefore, the wireless transmission latency for the EA to multicast its KV cache to the AAs is formulated as
\begin{align}
\label{eq:communication_latency_latent_space_e}
   t_{\text{KV,com},0} = \frac{m_{\text{KV},0}}{R_0}.
\end{align}

Subsequently, each AA de-quantizes the received bitstream into floating-point KV tensors, populates them into its local GPU memory, and concatenates the received KV cache with its pre-existing local cache.
For each layer of the Transformer, the KV matrices are reconstructed as $\mathbf{K}, \, \mathbf{V} \in \mathbb{R}^{(s_{0,1} + \alpha_{0,1}) \times (H d_h)}$.
Leveraging these inherited contextual states, each AA directly initiates the autoregressive decoding process, whose latency is derived by
\begin{align}
    \label{eq:AA_KV_inference_latency1}
    t_{\text{KV,inf},i,1} = t_{\text{NL,au},i,1}(\alpha_{i,1}, s_{0,1} + \alpha_{0,1}), \quad i \in \{1, \cdots, I\},
\end{align}
Thereafter, in the second round, every AA performs a full inference process by inputting new data, during which the incurred latency is expressed as
\begin{align}
    \label{eq:AA_KV_inference_latency2}
    t_{\text{KV,inf},i,2} &= t_{\text{NL,pr},i,2} (s_{i,2}, \omega_i) + t_{\text{NL,au},i,2} (\alpha_{i,2}, \omega_i ), \quad i \in \{1, \cdots, I\}, 
\end{align}
where $\omega_i = (s_{0,1} + \alpha_{0,1}) + \alpha_{i,1} + s_{i,2}$ {represents the historical context length}.

Subsequently, each AA transmits the newly generated KV cache back to the EA. 
The bit volume transmitted back to the EA from the $i$-th AA and the corresponding latency are expressed as
\begin{align}
    \label{eq:KV_cache_size2}
    &m_{\text{KV},i} = 32 L H d_h \frac{ (\alpha_{i,1} + s_{i,2}) + \alpha_{i,2}}{\gamma_i}, \quad i \in \{1, \cdots, I\},\\
    \label{eq:communication_latency_latent_space_w}
    &t_{\text{KV,com},i} = \frac{m_{\text{KV},i}}{R_i}, \quad i \in \{1, \cdots, I\}.
\end{align}

The EA de-quantizes and integrates the received KV caches into its local memory, thereby inheriting the collective insights of all AAs. 
Then, the EA embarks on the final autoregressive decoding phase, whose latency is given by
\begin{align}
    \label{eq:EA_KV_inference_latency2}
    t_{\text{KV,inf},0,2} &= t_{\text{NL,au},0,2} (\alpha_{0,2}, \varepsilon),
\end{align}
where $\varepsilon = (s_{0,1} + \alpha_{0,1}) + \sum_{i=1}^{I} \left[ \alpha_{i,1} + (s_{i,2} + \alpha_{i,2}) \right]$ {represents the total historical context length in the EA before initiating the final autoregressive phase.
It is noteworthy that $\omega_i$ and $\varepsilon$ are task-dependent exogenous parameters. They affect performance but are beyond our active control.}

Finally, we obtain the total E2E latency of the KV mode as
\begin{align}
    t_{\text{KV}} &= \sum_{j = 1}^{2} t_{\text{KV,inf},0,j} +t_{\text{KV,com},0} + \max_{i \in \{1,2, \cdots, I \} } \left\{ \sum_{j=1}^{2} t_{\text{KV,inf},i,j} +t_{\text{KV,com},i} \right\}.
    \label{eq:total_latency_latent_space}
\end{align}





\subsection{Comparison and Discussion}
\label{sec:comparison_and_discussion}
To compare the E2E latency performance of the two modes in the proposed scenario, we first establish the following conventions for analytical tractability.
Without loss of generality, let $i=1$ be the index that maximizes the AA related term.
Moreover, for simplicity, we assume $C_i = C$, and $s_{0,1} = s_{i,2} = s$ for any $i \in \{1, \cdots, I\}$.
Considering that the output quantity is regulable compared to the uncontrollable input size, we let $\alpha_{i,j} = \beta_{i,j} s_{i,j}$, with the exception of the final decision output which is fixed at $\alpha_{0,2} = 100$ due to its typically concise nature. 
For the subsequent numerical analysis, we further simplify the model by letting $\beta_{i,j} = \beta$ across all agents and turns.
Notably, $\beta$ varies across different inference tasks. For instance, $\beta \approx 1$ in translation tasks, whereas $\beta < 1$ in summarization tasks.
Additionally, we assume the channel conditions are represented by the signal-to-noise ratio (SNR), such that $ \frac{P_0 h_{\text{EA},i}^2}{B N_0} = \frac{P_i h_{\text{AE},i}^2} {\rho_i B N_0} = \text{SNR} $, and the bandwidth allocation ratio is uniformly distributed, i.e., $\rho_i = 1/I$.

\begin{table}[!h]
    \footnotesize
    \caption{Simulation Parameters Configuration}
    \label{tab:sim_params}
    \tabcolsep 0pt 

    \begin{tabular*}{\textwidth}{@{\hspace{2cm}} l @{\extracolsep{\fill}} l l @{\hspace{2cm}}}
        \toprule
        \textbf{Parameter Description} & \textbf{Symbol} & \textbf{Value} \\
        \midrule

        \multicolumn{3}{c}{\emph{LLM Parameters (based on LLaMA-7B~\cite{LLaMA2023_arXiv})}} \\ \midrule
        Transformer layer number               & $L$                                            & 32 \\
        Attention heads number                 & $H$                                            & 32 \\
        Head dimension                         & $d_h$                                          & 128 \\
        Hidden dimension                       & $d_m$                                          & 4096 \\
        FFN expansion dimension                & $d_f$                                          & 11008 \\
        Vocabulary size                        & $d_v$                                          & 32000 \\
        \midrule

        \multicolumn{3}{c}{\emph{Communication and Computation Parameters}} \\ \midrule
        Number of AAs                          & $I$                                            & 5 \\   
        Bits per token                         & $b$                                            & 16 \\
        KV cache compression ratio             & $\gamma_0, \gamma_i, i\in\{1,2, \cdots, I\}$   & 2 \\
        Transmission bandwidth                 & $B$                                            & 2 GHz \\
        Bandwidth allocation ratio             & $\rho_i, i\in\{1,2, \cdots, I\}$               & $1/I$ \\
        Computing capability                   & $C_0, C_i, i\in\{1,2, \cdots, I\}$             & 10 TFLOPS \\
        \midrule

        \multicolumn{3}{c}{\emph{Task Specific Parameters}} \\ \midrule
        Output/Input token ratio       & $\beta$         & 1 \\
        \bottomrule
    \end{tabular*}
\end{table}

\begin{figure}[!h]
    \centering

    \subfloat[]{
        \label{fig:beta}
        \includegraphics[width=0.45\textwidth]{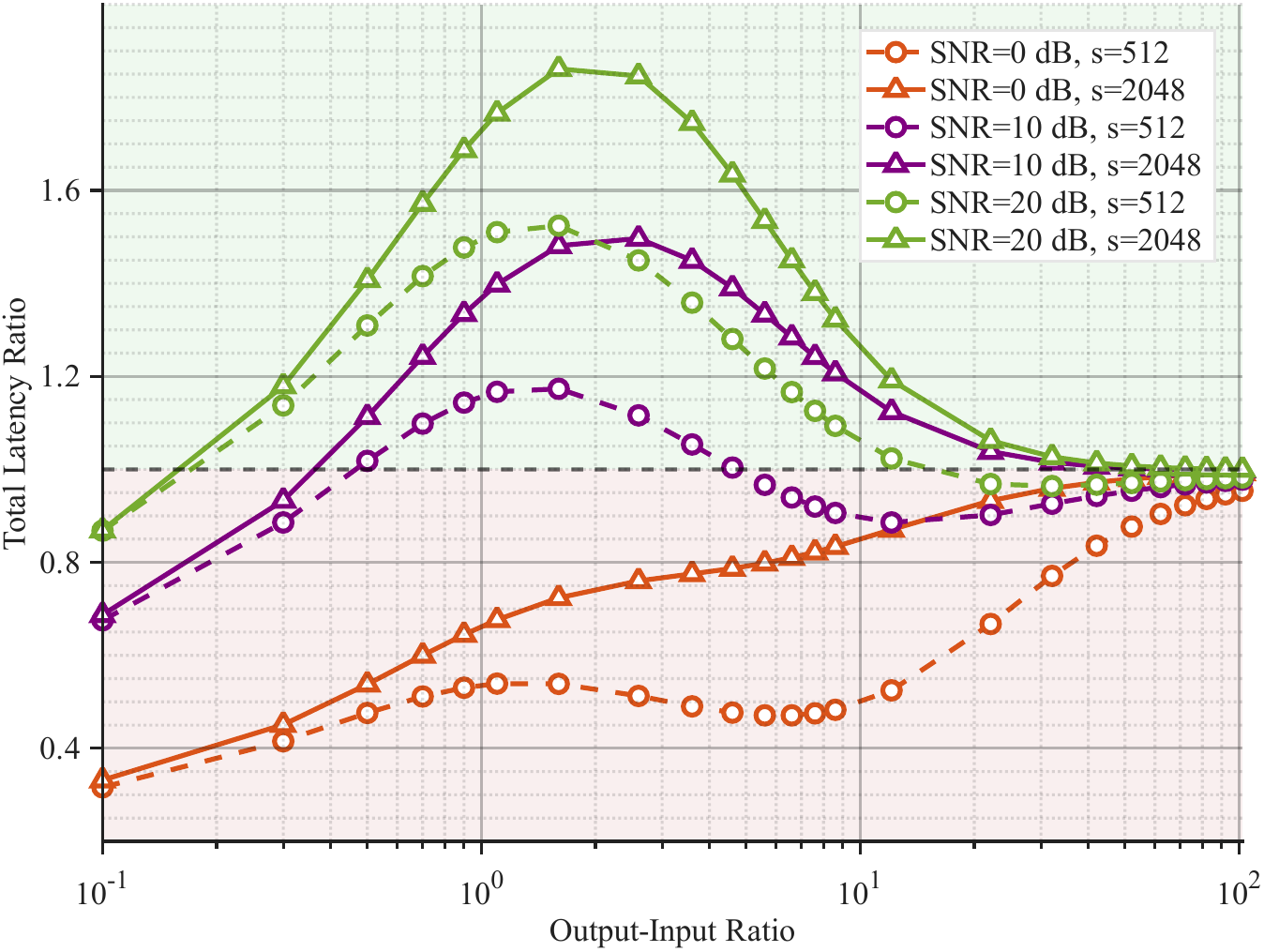}
    }
    \hfil
    \subfloat[]{
       \label{fig:compu_ratio} 
        \includegraphics[width=0.45\textwidth]{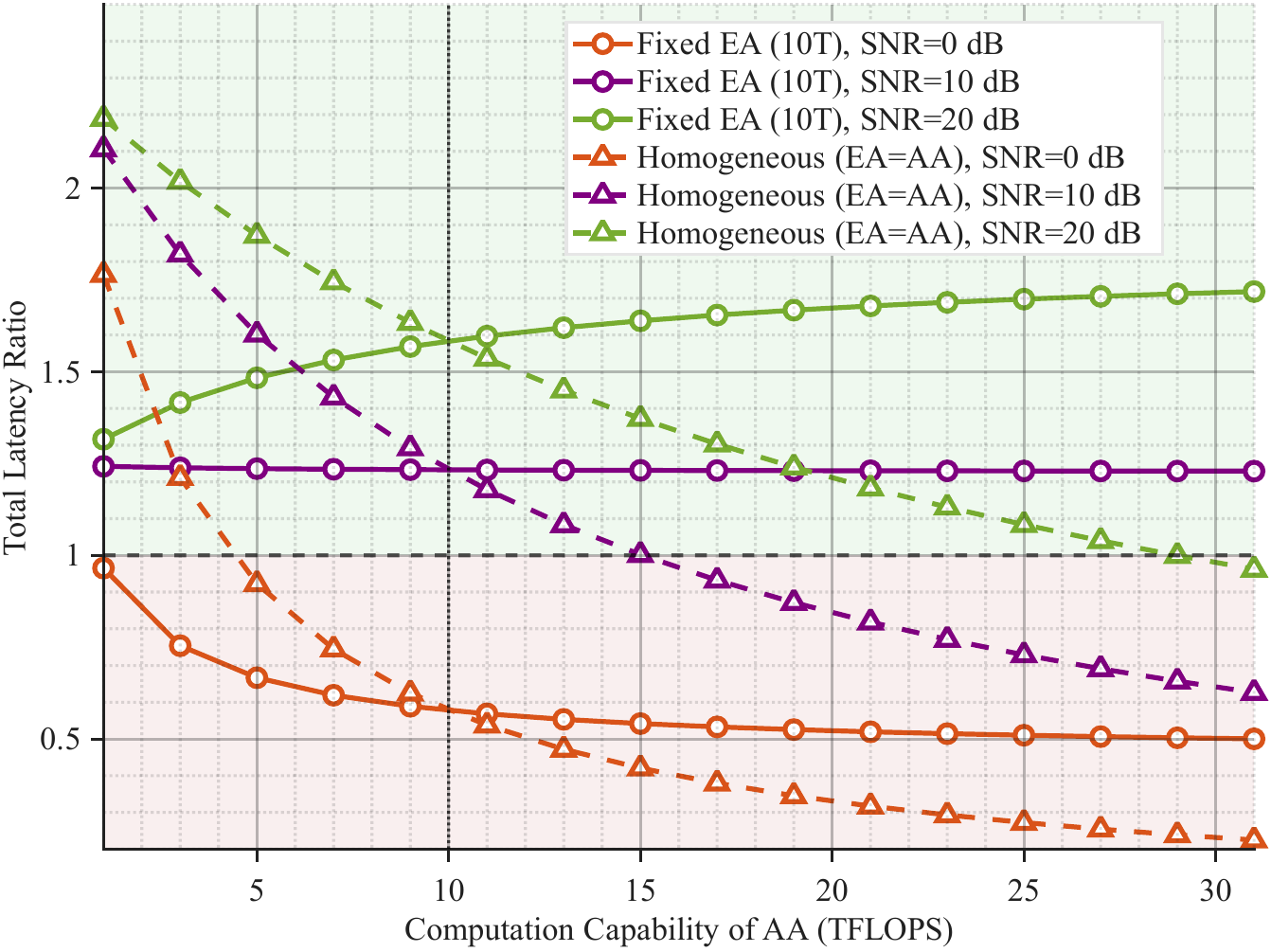}
        
    }
    \\
    \subfloat[]{
        \label{fig:snr} 
        \includegraphics[width=0.45\textwidth]{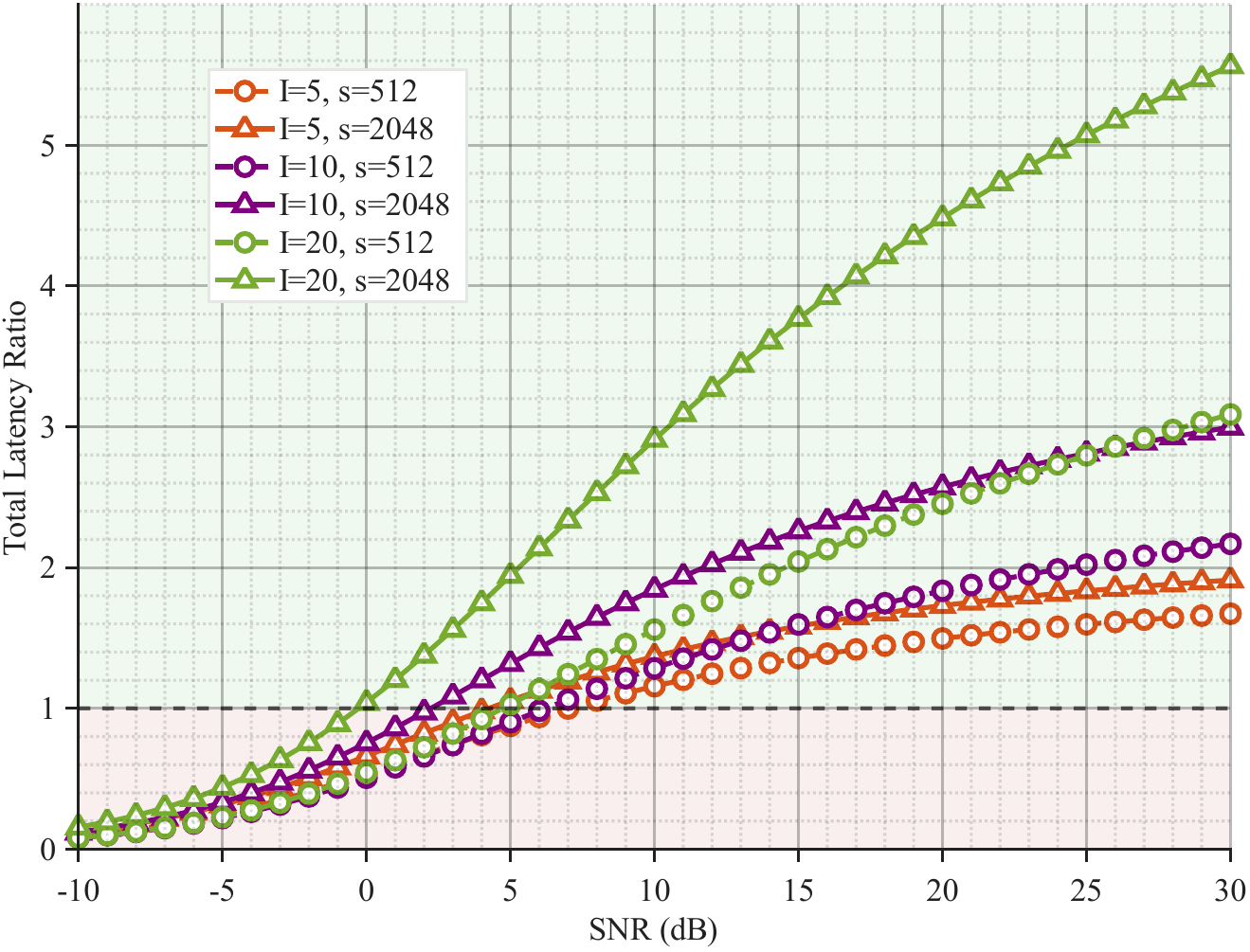}
    }
    \hfil
    \subfloat[]{
        \label{fig:I}
        \includegraphics[width=0.45\textwidth]{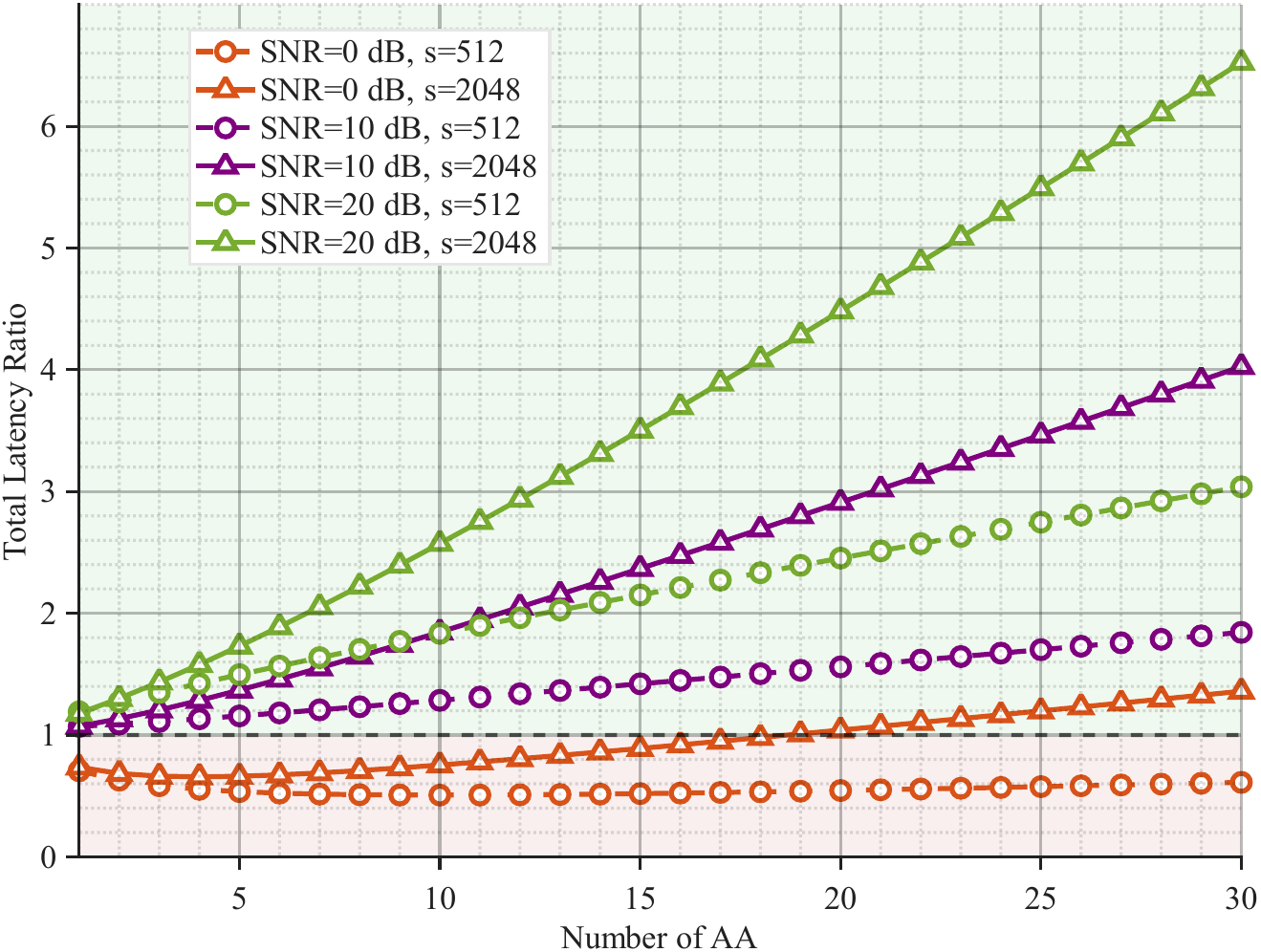}
    }

    \caption{E2E latency ratio $t_{\text{NL}}/t_{\text{KV}}$ versus (a) the output-input length ratio $\beta$, (b) the computation capacity $C_i$, (c) SNR of wireless channel, and (d) the number of AA $I$.}
    \label{fig:comparison} 
\end{figure}

The default simulation parameters are listed in \cref{tab:sim_params} and remain constant throughout the evaluation unless stated otherwise. 
Under these settings, we evaluate the relative E2E latency performance of the two modes by analyzing $t_{\text{NL}}/t_{\text{KV}}$, as illustrated in \cref{fig:comparison}. 
Specifically, a ratio below unity highlights the advantage of the NL mode (red area), whereas a ratio exceeding 1 implies the KV mode is more effective (green area).

\cref{fig:comparison} comprehensively illustrates the relative E2E latency performance ($t_{\text{NL}}/t_{\text{KV}}$) across distinct dimensions, i.e., task characteristics, device capabilities, and network conditions.
We can observe a common theme across all subfigures: neither the NL nor the KV mode maintains a consistent dominance. 
Instead, the optimal mode shifts dynamically based on the system configuration.
Specifically, the results reveal the following trade-offs.

\emph{1) Task Dependency (\cref{fig:comparison} (a)): }
In fixed-scale generation tasks, while the performance gap between NL and KV modes narrows with an increasing $\beta$, the underlying trend is non-monotonic, rendering the optimal transmission mode elusive and difficult to predict. 
Additionally, the result is highly sensitive to the varying SNR.

\emph{2) Hardware Heterogeneity (\cref{fig:comparison} (b)): }
The latency performance is highly sensitive to the computing capabilities of the EA and AAs.
However, we observe that, as the EA's computing capability increases, the optimal mode gradually converges towards the NL mode. 
This is attributed to the fact that the enhanced EA capability reduces the computation latency during the prefill phase when receiving all the data from AAs.

\emph{3) Channel Sensitivity (\cref{fig:comparison} (c)): }
A monotonic increase in the latency ratio is observed with improving SNR. 
High-quality channels facilitate the data-intensive KV mode, while poor channels force the MAS change to the NL mode to avoid transmission delays.

\emph{4) Scale Impact (\cref{fig:comparison} (d)): }
The latency ratio rises with the number of AA. 
Interestingly, a denser network tends to favor the KV mode, suggesting that the computing overhead in the NL mode scales less flexibility than the transmission in the KV mode.

In summary, the ``crossover point'' (i.e., $t_{\text{NL}}/t_{\text{KV}} = 1$) exists in all four subfigures and shifts significantly as parameters change.
This phenomenon demonstrates that a static mode selection strategy is insufficient to handle the volatility of embodied scenarios.
Consequently in order to minimize the E2E latency in this scenario as much as possible, it is imperative to develop a scheme that can adaptively determine the best decision and allocate communication resources in response to the instantaneous task requirements ($s, \alpha$), device states ($C_0, C_i$), network environment ($\text{SNR}, I$) and other factors.


\section{Optimal Mode Selection under Practical Conditions}
\label{sec:optimal_mode_selection}
The performance analysis presented in \cref{sec:comparison_and_discussion} assumed an idealized environment where hardware and wireless channels remained static, with the system and environmental parameters simplified as being identical across all agents.
While this macroscopic perspective successfully demonstrates the computation-communication trade-off, practical MAS deployments operate under highly dynamic and heterogeneous conditions. 
In real-world embodied operations, not only do hardware capacities and wireless channels fluctuate, but the multi-turn conversational process itself is intrinsically stochastic.
Therefore, this section further investigates how to choose the best mode under variable practical conditions, especially the limited communication bandwidth.

\subsection{Bandwidth Constraint}
In A2A communication scenarios, bandwidth allocation is strictly governed by pre-established communication protocols rather than being utilized arbitrarily. 
Accordingly, in the context of embodied MASs, the available spectrum for all agents is constrained within a fixed total bandwidth $B$. 
This total bandwidth is shared among all agents, where the specific allocation strategy directly dictates the communication efficiency. 
For the first round conversation, the EA broadcasts to all AAs using the entire bandwidth $B$.
In the second round, each AA communicates with the EA with allocated bandwidth, denoted as $\rho_i B$, where $\rho_i$ represents the bandwidth allocation coefficient for the $i$-th AA.
The bandwidth assigned to each AA also determines the corresponding noise power, which in turn influences the SNR. 
Besides, as transmit power of each embodied agent varies and is typically limited by hardware capabilities and energy constraints, along with fluctuating channel conditions, the achievable data rate is directly affected by the allocated bandwidth.
Therefore, the bandwidth fraction allocated to each AA must be meticulously managed to ensure optimal communication performance.


\subsection{Mode Selection Analysis}
\label{sec:mode_selection_analysis}
Our analysis in \cref{sec:comparison_and_discussion} reveals that the NL mode is less demanding of environmental resources in the proposed scenario. 
Specifically, it is advantageous in scenarios with low SNR or low data transmission requirement.
However, if the computation capacity is limited, the NL mode may not be suitable.
This indicates that the optimal communication mode is contingent upon the prevailing resource conditions.
Furthermore, while the previous analysis in \cref{sec:latency_analysis} assumed a uniform mode across all agents, a practical MAS is inherently heterogeneous.
Different agents may employ different modes-some sending tokens and others sending KV caches-based on their specific local conditions.
In what follows, we investigate how to identify the optimal communication mode for each agent by evaluating the interplay among these dynamic factors.


\subsubsection{{Marginal Latency}}
\label{sec:marginal_latency}
To identify the optimal {mode for each specific transmission, we must refine the latency analysis. 
A key distinction from the global perspective adopted in \cref{sec:latency_analysis} is the transition to a marginal latency perspective based on a transmitting agent's local field of view. 
Specifically, a transmitting agent faces two fundamental constraints:
it can neither predict the receiver's subsequent autoregressive generation length (an unpredictable future) nor alter the receiver's accumulated historical context (an immutable sunk state). 
As the dialogue deepens, this historical context continuously accumulates. 
Due to the nature of the attention mechanism, the historical context acts as a dynamic exogenous parameter that heavily influences the computation workload.
Consequently, a realistic approach is to evaluate the system based on the immediate marginal pressure caused by the current data chunk under the given historical state. 
This formulation makes our mode selection strategy naturally applicable to conversations of arbitrary length.
It is worth noting that when extending the dialogue from two rounds to an arbitrary number of turns, the interaction still follows a sequential turn-taking paradigm, which is consistent with the signal flow described in \cref{sec:hierarchical_architecture}. 
Specifically, the EA must wait for the AA's response before initiating the next round, ensuring no overlapping transmissions.
More importantly, this evaluation applies to any transmitting agent and any receiving agent in the system, denoted by $i_t$ and $i_r$ respectively, where $i_t, i_r = 0$ represents the EA while $i_t, i_r \neq 0$ represents an AA.}

For a transmitter agent employing token-based transmission, the marginal latency resulting from its current $\alpha_{i_t,j}$ output tokens is given by
\begin{align}
    \label{eq:delta_latentcy_natural_language}
    t^{\triangle}_{\text{NL},i_t,i_r,j} (\alpha_{i_t,j}) = \underbrace{t_{\text{NL,pr},i_r,j} (\alpha_{i_t,j}, \alpha_{i_t,j} + {\theta_{i_r,j}})}_{\text{{prefill latency at the receiver}}} + \underbrace{{b \alpha_{i_t,j} / R_{i_t,j}}}_{\text{{wireless transmission latency}}}, \quad i_t, i_r \in \{0, 1, \cdots, I \}, \quad i_t \ne i_r,
\end{align}
{where the first term on the right side of the equals sign comes from \cref{eq:inference_latency_natural_language1} and the second term (similar to \cref{eq:communication_latency_natural_language_e,eq:communication_latency_natural_language_w}) denotes the wireless transmission latency under current channel capacity $R_{i_t,j}$.
If $i_t = 0$, $R_{i_t,j}$ refers to \cref{eq:R0}.
Otherwise, $R_{i_t,j}$ is instructed by \cref{eq:Ri}.}
{
As shown in \cref{eq:delta_latentcy_natural_language}, the inputs of the prefill latency function are solely $\alpha_{i_t,j}$.
This is because the receiver's prefill input length $s_{i,j}$ exactly equals the transmitter's output token volume $\alpha_{i_t,j}$.}
{Apart from the self-attention computed among the input tokens themselves, they must also attend to the historical context of length $\theta_{i_r,j}$, introducing a computational overhead that continuously accumulates as the dialogue deepens.}

Conversely, if the transmitter agent opts for KV cache transmission, it knows the receiver will bypass the prefill stage to save computational latency.
Thus, the corresponding marginal latency is given by
\begin{align}
    \label{eq:delta_latency_kv}
    t^{\triangle}_{\text{KV},i_t,i_r,j} (\alpha_{i_t,j}) = \underbrace{{ \frac{32 L H d_h ( \xi_{i_t,j} + \alpha_{i_t,j} ) / \gamma_{i_t,j}}{R_{i_t,j}} }}_{\text{{wireless transmission latency}}}, \quad i_t, i_r \in \{0, 1, \cdots, I \}, \quad i_t \ne i_r,
\end{align}
{where the expression in the numerator represents the transmission volume (in bits) caused by the KV cache, which is consistent with the derivations in \cref{eq:KV_cache_size1,eq:KV_cache_size2}.
\cref{eq:delta_latency_kv} is similar to but distinct from \cref{eq:communication_latency_latent_space_e,eq:communication_latency_latent_space_w}. 
Under dynamic mode switching, the transmission volume fluctuates based on wireless channel conditions and conversation parameters, rendering the static formulations in \cref{eq:communication_latency_latent_space_e,eq:communication_latency_latent_space_w} inapplicable. 
To mitigate the impact of the wireless environment on KV semantics, the quantization precision $\gamma_{i_t,j}$ should be dynamically adapted for each transmission.
Furthermore, to avoid redundant data transfers, the transmitter sends only the KV cache blocks absent from the receiver's memory. 
Therefore, we introduce $\xi_{i_t,j}$ to represent the context dimension of the unshared historical KV cache for transmitter $i_t$ at round $j$.
$\xi_{i_t,j}$ is fluctuating but can be derived from previous interactions.
For instance, $\xi_{0,1} = s_{0,1}$ when $j=1$, $\xi_{i_t,2} = \alpha_{i_t,2} + s_{i_t,2}$ or $\xi_{i_t,2} = s_{i_t,1} + \alpha_{i_t,2} + s_{i_t,2}$ when $j=2$.}

{
\textbf{Remark:} As the multi-agent dialogue deepens, a practical challenge emerges regarding the unshared historical dimension $\xi_{i_t,j}$. 
If the agents predominantly communicate via the token mode for numerous consecutive turns, the unshared state accumulates rapidly. 
Consequently, the rapid accumulation of $\xi_{i_t,j}$ would render the KV mode permanently suboptimal due to prohibitive wireless delays, ultimately locking the system into the token mode and paralyzing the mode-switching mechanism.
To prevent infinite accumulation, the practical KV transmission volume must be upper-bounded, which is justified by the following constraints:
1) \textbf{Context Window Limits:} Practical LLMs have a maximum context window length. Both the transmitter and the receiver will inevitably truncate their oldest KV caches, meaning that $\xi_{i_t,j}$ and $\theta_{i_r,j}$ cannot grow indefinitely.
2) \textbf{Memory-Aware Transmission:} This ``memory-awareness" is dual-faceted. First, it respects the physical memory constraints, ensuring the incoming KV payload does not exceed the receiver's available hardware capacity.  Second, it preserves the interactive memory, preventing an oversized transfer from forcing the receiver to evict its local historical context to accommodate new tensors. Such an overwrite would sever dialogue continuity, erase the receiver's cognitive history, and undermine the purpose of multi-agent collaboration.
}


\subsubsection{{Single-Link Mode Selection}}
\label{sec:single_link_mode_selection}
After the aforementioned discussion, to determine the best transmission mode, we now proceed to formulate the mode selection criterion for a point-to-point link.

Let $f_{i_t,i_r,j} = t^{\triangle}_{\text{NL},i_t,i_r,j} (\alpha_{i_t,j}) - t^{\triangle}_{\text{KV},i_t,i_r,j} (\alpha_{i_t,j})$, and then we obtain
\begin{align}
    \label{eq:f}
    f_{i_t,{i_r},j} (\alpha_{i_t,j}) = k_4 \alpha^2_{i_t,j} + k_5 \alpha_{i_t,j} + k_6, \quad i_t, i_r \in \{0, 1, \cdots, I \}, \quad i_t \ne i_r,
\end{align}
where $k_4 = \frac{2 k_1}{C_{i_r}}$, $k_5 = \frac{k_2 + {2 k_1 \theta_{i_r,j}}}{C_{i_r}} + \frac{b - 16 k_1/ \gamma_{i_t,{j}}}{R_{i_t,{j}}}$, $k_6 = \frac{k_3}{C_{i_r}} - \frac{16 k_1 \xi_{i_t,{j}} / \gamma_{i_t,{j}}}{R_{i_t,{j}}}$.
It is evident that $f_{i_t,i_r,j} (\alpha_{i_t,j})$ is a quadratic function of $\alpha_{i_t,j}$.
The sign of $k_4$ is always positive, indicating that the parabola opens upwards.
The superior communication mode can be determined simply by evaluating the objective function at the current value of $\alpha_{i_t,j}$ and other system parameters.
If $f_{i_t,i_r,j} (\alpha_{i_t,j}) > 0$, the KV mode is preferred.
Otherwise, the NL mode is more advantageous.
A close examination of \cref{eq:f} reveals that the optimal selection between the NL and KV modes is dictated by a multifaceted interplay of several factors. 
These include the agents' computational capacities, wireless transmission rates (which are functions of transmit power, channel gains, and bandwidth allocation coefficient), and dynamic conversation cost. 

To this end, we further analyze the coupling between the bandwidth allocation and the optimal mode selection.
\cref{eq:f} can be rewritten as a function of the bandwidth allocation coefficient $\rho_{i_t,j}$, expressed as
\begin{align}
    \label{eq:f(rho)}
    f_{i_t,{i_r,j}} (\alpha_{i_t,{j}}, \rho_{i_t,{j}}) = A(\alpha_{i_t,{j}}) - \frac{D(\alpha_{i_t,{j}})}{R(\rho_{i_t,{j}})}, \quad {i_t, i_r \in \{0, 1, \cdots, I \}, \quad i_t \ne i_r},
\end{align}
where $A(\alpha_{i_t,{j}}) = [2 k_1 \alpha^2_{i_t,{j}} + (k_2+{2k_1\theta_{i_r,j}}) \alpha_{i_t,{j}} + k_3 ]  / C_{{i_r}}$, $D(\alpha_{i_t,{j}}) = 16 k_1 (\xi_{i_t,{j}} + \alpha_{i_t,{j}}) / \gamma_{i_t,{j}}  - b \alpha_{i_t,{j}}$, and $R(\rho_{i_t,{j}}) = \rho_{i_t,{j}} B \log_2 (1 + \frac{P_{i_t,{j}} h^2_{i_t,{j}}/B N_0}{\rho_{i_t,{j}}})$. 
We observe that $D(\alpha_{i_t,j}) > 0$ generally holds since $k_1$ is significantly large, and $R(\rho_{i_t,j})$ is a monotonically increasing function of $\rho_{i_t,j}$.
Consequently, $f_{i_t,i_r,j}(\alpha_{i_t,j}, \rho_{i_t,j})$ is monotonically increasing with respect to $\rho_{i_t,j}$.
This implies the existence of a unique bandwidth threshold, denoted as $\rho_{i_t,j}^*$, which satisfies $f_{i_t,i_r,j} (\alpha_{i_t,j}, \rho^*_{i_t,j}) = 0$. 
Thus, KV mode is advantageous when $\rho_{i_t,j} > \rho_{i_t,j}^*$.
Otherwise, the NL mode is better.
\begin{figure}[!t]
    \centering
    \includegraphics[width=0.98\textwidth]{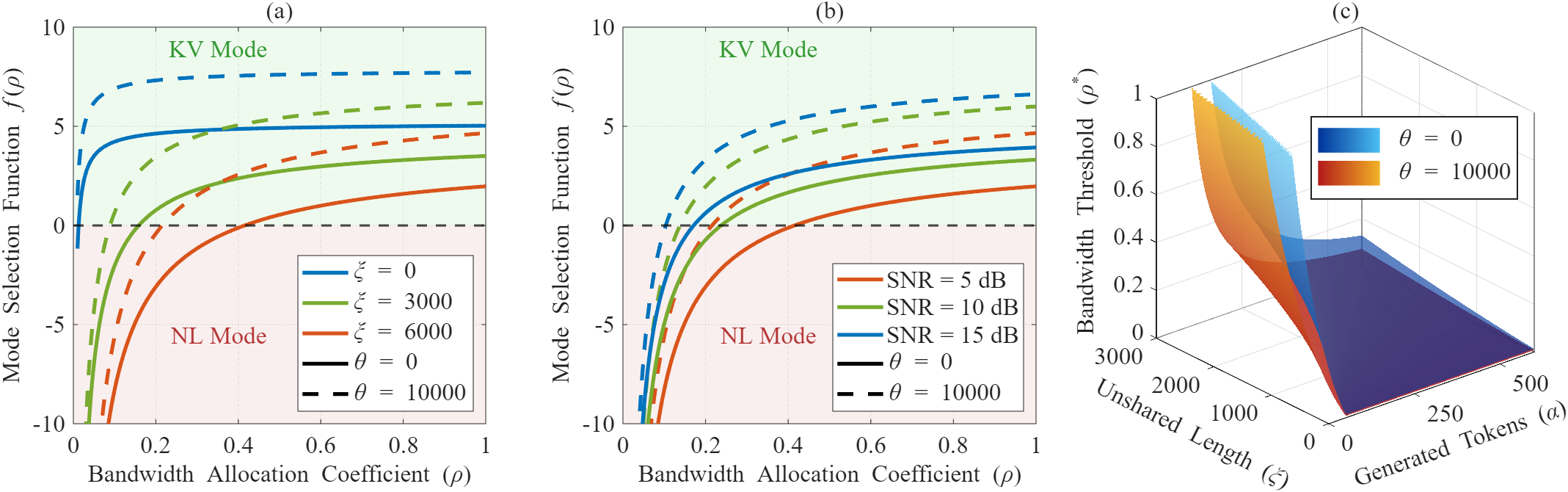} 
    \caption{Visualization of the single-link mode selection criterion. (a) Impact of unshared context length $\xi$ (SNR = 5 dB). (b) Impact of wireless channel quality SNR ($\xi = 6000$). (c) 3D decision boundary of the bandwidth allocation coefficient $\rho^*$ (SNR = 5 dB).}
    \label{fig:performance_case_study_signle_link} 
\end{figure}

{
Although \cref{eq:f(rho)} is formulated for a point-to-point link, it is inherently applicable to the EA-to-AAs broadcast scenario. 
Given that the EA exclusively monopolizes the bandwidth, and to ensure reliable decoding without exceeding any agent's Shannon capacity, the EA's broadcast rate is strictly dictated by the worst-case channel condition. 
This implies that all AAs effectively experience the exact same bottleneck channel, stripping away the channel heterogeneity. 
As a result, evaluating the broadcast latency is mathematically equivalent to assessing $I$ independent single links.
Thus, the best mode can be found via a min-max criterion among these $I$ links.}

{
To visually corroborate the monotonic behavior and the existence of the decision boundary $\rho^*$ derived in ~\cref{eq:f(rho)}, \cref{fig:performance_case_study_signle_link} illustrates the objective function $f(\rho)$ and the 3D threshold surface. 
For visualization purposes, we assume a baseline setting with receiver computation capacity $C_{i_r} = 1$ TFLOPS, generated tokens $\alpha = 512$, KV cache compression ratio $\gamma = 2$, wireless bandwidth $B = 2$ GHz, and noise power spectral density $N_0 = -140$ dBm/Hz. 
Note that here $\rho B$ denotes the effective bandwidth allocated to the transmitter.
As depicted in \cref{fig:performance_case_study_signle_link} (a) and (b), $f(\rho)$ strictly increases with $\rho$. 
As $f(\rho) \to -\infty$ when $\rho \to 0$, the existence of $\rho^*$ can be ascertained simply by evaluating if $f(1) > 0$. If it exists, $\rho^*$ can be efficiently obtained using the bisection method.
Moreover, a heavier unshared context length $\xi$ or a poorer wireless channel pushes the threshold $\rho^*$ to higher values, thereby making the NL mode more attractive.
Besides, a larger historical context $\theta$ at the receiver elevates the computational overhead of the NL mode, which in turn decreases $\rho^*$ and makes the transmitter favor the KV mode.
\cref{fig:performance_case_study_signle_link} (c) further reveals the 3D decision boundary. 
Observed along the $\alpha$ axis, as $\alpha$ increases, $\rho^*$ decreases, prompting the transmitter to prefer the KV mode. 
Conversely, accumulating historical KV debt (increasing $\xi$) imposes a higher bandwidth requirement for the KV mode.
}

\subsubsection{{Joint Mode and Bandwidth Optimization}}
\label{sec:joint_optimization}
Unlike the contention-free broadcast scenario, the returning links from the AAs to the EA introduce severe wireless resource contention. 
Because the total shared bandwidth is strictly limited, any bandwidth acquired by one agent inherently comes at the expense of the others. 
Consequently, the point-to-point criterion in \cref{eq:f(rho)} is inapplicable to this scenario. 
Specifically, enforcing a greedy mode selection based on isolated, pre-calculated thresholds (where each agent acts solely on its individual requirement) may disproportionately monopolize the bandwidth pool, leading to suboptimal global performance or even system infeasibility. 
Therefore, rather than treating mode selection as a passive outcome of a predetermined bandwidth allocation, we further establish a joint optimization framework to holistically minimize the marginal latency. 
To this end, we define the continuous bandwidth allocation vector $\boldsymbol{\rho} \triangleq [\rho_1, \dots, \rho_I]^T$ and the binary mode selection vector $\boldsymbol{x} \triangleq [x_1, \dots, x_I]^T$, where $x_i = 0$ indicates the NL mode and $x_i = 1$ indicates the KV mode.
{For simplicity, we omit the round index $j$ in the following formulations.}

Unlike threshold-based method, we allow the MAS to flexibly optimize $\boldsymbol{x}$ and $\boldsymbol{\rho}$ to balance the trade-off between computational and communication overheads globally.
Consequently, we formulate a latency minimization optimization problem as
\begin{align} 
    (\mathcal{P}1) \quad \min_{\boldsymbol{x}, \boldsymbol{\rho}} \quad & J = \underbrace{t_{\text{cmp},0} \left( \boldsymbol{x}, \boldsymbol{\alpha} \right)}_{{\text{EA prefill latency}}}  + \max_{i} \{ \underbrace{t_{\text{com},i} (x_i, \rho_i) }_{{\text{AA wireless transmission latency}}} \} \\
    \text{s.t.} \quad & \sum_{i=1}^{I} \rho_i \leqslant 1, \quad \rho_i > 0, \quad \forall i \in \{1, \cdots, I\}, \\
                      & x_i \in \{0, 1\}, \quad \forall i \in \{1, \cdots, I\}, 
\end{align}
where $\boldsymbol{\alpha} \triangleq [ \alpha_{1}, \alpha_{2}, \cdots, \alpha_{I} ]^T$ is output token count vector for all AAs.
$t_{\text{cmp},0} (\boldsymbol{x}, \boldsymbol{\alpha}) = t_{\text{NL,pr},0,j} ((\boldsymbol{1}-\boldsymbol{x})^T \boldsymbol{\alpha}, (\boldsymbol{1}-\boldsymbol{x})^T \boldsymbol{\alpha} + {\theta_{0,j}})$ is the prefill latency at the EA.
$t_{\text{com},i} (x_i, \rho_i) = \frac{D_i(x_i)}{R(\rho_i)}$ is the wireless transmission latency of the $i$-th AA, where $D_i(x_i) = (1-x_i) D_{i, \text{NL}} + x_i D_{i, \text{KV}}$ represents the transmitted data volume, $D_{i, \text{NL}}$ and $D_{i, \text{KV}}$ are the NL data volume and KV cache volume respectively.
{Note that, as discussed in \cref{sec:marginal_latency}, $D_{i,\text{NL}}$ is determined by $\alpha_i$ while $D_{i,\text{KV}}$ is determined by $\alpha_i$ and $\xi_i$.
Moreover, while asynchronous transmission could theoretically improve bandwidth utilization, the autoregressive inference latency of LLM-driven AAs is inherently task-dependent and unpredictable. 
To avoid predicting autoregressive latencies of AAs and satisfy the strict synchronization requirements of OFDMA, we assume all AAs initiate transmission simultaneously.
This synchronous protocol dictates the min-max structure of the formulation, thereby enabling the tractable joint optimization of mode selection and bandwidth allocation.}

\begin{algorithm}[!t]
    \caption{Joint Mode Selection and Resource Allocation Strategy (JMSRA)}
    \label{alg:jmsra}
    \begin{algorithmic}[1]
        \REQUIRE System parameters $\alpha_{i}, \xi_i, \gamma_i, P_i, h_{i}, C_0, \theta_{0}, I, B, N_0$, data volumes $D_{i, \text{NL}}, D_{i, \text{KV}}$, error tolerance $\delta$.
        \ENSURE Optimal mode vector $\boldsymbol{x}^{\star}$ and bandwidth allocation $\boldsymbol{\rho}^{\star}$.
        
        \STATE \textbf{Function} $\textsc{GreedySearch}(\boldsymbol{x}_{\text{init}}, \text{target\_val})$:
        \STATE \quad $\boldsymbol{x} \leftarrow \boldsymbol{x}_{\text{init}}$; \quad $\mathcal{S} \leftarrow \{i \mid x_i \neq \text{target\_val}\}$;
        \STATE \quad $J_{\text{curr}} \leftarrow \textsc{CalcLatency}(\boldsymbol{x})$;
        \STATE \quad \textbf{while} $\mathcal{S} \neq \emptyset$ \textbf{do}
        \STATE \quad \quad $k^{\star} \leftarrow \arg\max_{k \in \mathcal{S}} \{ J_{\text{curr}} - \textsc{CalcLatency}(\boldsymbol{x} \text{ with } x_k \leftarrow \text{target\_val}) \}$;
        \STATE \quad \quad $J_{\text{next}} \leftarrow \textsc{CalcLatency}(\boldsymbol{x} \text{ with } x_{k^{\star}} \leftarrow \text{target\_val})$;
        \STATE \quad \quad \textbf{if} $J_{\text{next}} < J_{\text{curr}}$ \textbf{then} $\boldsymbol{x}[k^{\star}] \leftarrow \text{target\_val}$; $J_{\text{curr}} \leftarrow J_{\text{next}}$; $\mathcal{S} \leftarrow \mathcal{S} \setminus \{k^{\star}\}$;
        \STATE \quad \quad \textbf{else} \textbf{break};
        \STATE \quad \textbf{return} $\boldsymbol{x}, J_{\text{curr}}$;
        {
        \STATE \textbf{Function} $\textsc{BisectionSearch}(\boldsymbol{x})$:
        \STATE \quad Initialize latency bounds: $T_{\text{low}} \leftarrow \epsilon$, $T_{\text{high}} \leftarrow T_{\text{max}}$;
        \STATE \quad \textbf{while} $T_{\text{high}} - T_{\text{low}} > \delta$ \textbf{do}
        \STATE \quad \quad $T_{\text{try}} \leftarrow (T_{\text{low}} + T_{\text{high}}) / 2$; \quad $\rho_{\text{sum}} \leftarrow 0$;
        \STATE \quad \quad \textbf{for} $i = 1$ \textbf{to} $I$ \textbf{do}
        \STATE \quad \quad \quad $D_i \leftarrow (1-x_i)D_{i,\text{NL}} + x_i D_{i,\text{KV}}$;
        \STATE \quad \quad \quad Determine minimum bandwidth $\rho_i^{\text{req}}$ to satisfy $t_{\text{com},i}(\rho_i^{\text{req}}) \le T_{\text{try}}$;
        \STATE \quad \quad \quad $\rho_{\text{sum}} \leftarrow \rho_{\text{sum}} + \rho_i^{\text{req}}$;
        \STATE \quad \quad \textbf{if} $\rho_{\text{sum}} \le 1$ \textbf{then} $T_{\text{high}} \leftarrow T_{\text{try}}$; \COMMENT{Feasible}
        \STATE \quad \quad \textbf{else} $T_{\text{low}} \leftarrow T_{\text{try}}$; \COMMENT{Infeasible}
        \STATE \quad \textbf{return} $T_{\text{high}}$ (as $\tau^{\star}$), and corresponding allocation $\boldsymbol{\rho}^{\text{req}}$;}

        \STATE \textbf{Function} $\textsc{CalcLatency}(\boldsymbol{x})$:
        \STATE \quad $t_{\text{cmp},0}(\boldsymbol{x}) \leftarrow \text{Prefill latency at EA given } \boldsymbol{x}$;
        \STATE \quad $\tau^{\star} \leftarrow \textsc{BisectionSearch}(\boldsymbol{x})$;
        \STATE \quad \textbf{return} $t_{\text{cmp},0}(\boldsymbol{x}) + \tau^{\star}$;

        \STATE \textbf{Main Procedure:}
        \STATE $(\boldsymbol{x}_{\text{fwd}}, J_{\text{fwd}}) \leftarrow \textsc{GreedySearch}(\mathbf{0}, 1)$; \COMMENT{Forward: NL $\rightarrow$ KV}
        \STATE $(\boldsymbol{x}_{\text{bwd}}, J_{\text{bwd}}) \leftarrow \textsc{GreedySearch}(\mathbf{1}, 0)$; \COMMENT{Backward: NL $\leftarrow$ KV}
        \STATE \textbf{if} $J_{\text{fwd}} \le J_{\text{bwd}}$ \textbf{then} $\boldsymbol{x}^{\star} \leftarrow \boldsymbol{x}_{\text{fwd}}$ \textbf{else} $\boldsymbol{x}^{\star} \leftarrow \boldsymbol{x}_{\text{bwd}}$;
        \STATE $\boldsymbol{\rho}^{\star} \leftarrow \textsc{BisectionSearch}(\boldsymbol{x}^{\star})$; \COMMENT{Retrieve optimal bandwidth}
        \RETURN $\boldsymbol{x}^{\star}, \boldsymbol{\rho}^{\star}$;
    \end{algorithmic}
\end{algorithm}

One can see that ($\mathcal{P}1$) involves both binary variables $\boldsymbol{x}$ and continuous variables $\boldsymbol{\rho}$, making it a typical MINLP problem. 
The mathematical intractability of ($\mathcal{P}1$) is twofold.
First, the $\max(\cdot)$ operator renders the objective function non-smooth (i.e., non-differentiable).
Second, the coexistence of binary integer constraints and non-linear fractional terms makes the problem strictly non-convex. 
Generally, such MINLP problems are NP-hard. Relying on an exhaustive search for the binary vector $\boldsymbol{x}$ incurs a complexity of $\mathcal{O}(2^I)$, which imposes a prohibitive computational burden as the number of AAs increases. 
Therefore, an efficient algorithm with low complexity is required.

To tackle the intractability, we observe that despite the coupling between $\boldsymbol{x}$ and $\boldsymbol{\rho}$, if $\boldsymbol{x}$ is fixed, the term $t_{\text{cmp},0}(\boldsymbol{x}, \boldsymbol{\alpha})$ becomes a constant. 
Consequently, the original problem degenerates into a bandwidth allocation sub-problem, which is expressed as
\begin{align} 
    (\mathcal{P}2) \quad \min_{\boldsymbol{\rho}} \quad &  \tau = \max_{i} \{ t_{\text{com},i} (\rho_i) \} \\
    \text{s.t.} \quad & \sum_{i=1}^{I} \rho_i \leqslant 1, \quad \rho_i > 0, \quad \forall i \in \{1, \cdots, I\}.
\end{align}

For a given $\boldsymbol{x}$, the wireless transmission latency of the $i$-th AA is monotonically decreasing with respect to its allocated bandwidth $\rho_i$. 
This property implies that the optimal bandwidth allocation of ($\mathcal{P}2$) can be efficiently obtained using bisection search.
{Specifically, rather than directly searching over the multi-dimensional vector $\boldsymbol{\rho}$, the bisection search operates on a unified candidate latency value for all AAs, denoted by $T_{\text{try}}$. 
Given the monotonic strict decreasing property of $t_{\text{com},i}(\rho_i)$, for any trial latency $T_{\text{try}}$, the minimum required bandwidth $\rho_i^{\text{req}}$ for the $i$-th agent to satisfy $t_{\text{com},i}(\rho_i^{\text{req}}) \le T_{\text{try}}$ can be uniquely determined via the inverse function of $t_{\text{com},i}(\rho_i)$ or another bisection method. 
The feasibility of the trial latency $T_{\text{try}}$ is then evaluated by checking the sum bandwidth constraint $\sum_{i=1}^I \rho_i^{\text{req}} \le 1$. 
If the constraint is satisfied, $T_{\text{try}}$ is updated as the new upper bound to further push for a lower latency; otherwise, it becomes the new lower bound. 
This process iterates until the latency bounds reach a predefined tolerance, yielding both the optimal latency and the corresponding optimal bandwidth allocation.
}

Since $\boldsymbol{\rho}$ can always be dynamically optimized to match any selected $\boldsymbol{x}$, the core challenge shifts to finding the optimal $\boldsymbol{x}$ without resorting to a exhaustive search across the $2^I$ state space. 
This motivates us to design a joint optimization framework that iteratively updates $\boldsymbol{x}$ and optimizes $\boldsymbol{\rho}$ accordingly.
Instead of relying on a random or exhaustive search, we propose to explore the solution space of $\boldsymbol{x}$ in a greedy manner. 
Specifically, by evaluating the marginal latency reduction of switching the transmission mode of specific agents, we can progressively approach the optimal solution.
To further improve the robustness and mitigate the risk of local optima, we introduce a bidirectional search strategy.
In summary, we propose the JMSRA algorithm, summarized in \cref{alg:jmsra}. 
The details are as follows.

As illustrated in \cref{alg:jmsra}, the algorithm operates in a nested architecture: an outer loop for discrete mode selection and an inner kernel for continuous bandwidth allocation.

\emph{1) Outer Loop (Bidirectional Greedy Search):} 
To circumvent the prohibitive complexity of exhaustive search, we employ a bidirectional greedy heuristic.
The forward search initializes with the ``All-NL" state ($\boldsymbol{x} = \mathbf{0}$, computation-intensive for the EA) and iteratively switches the AA offers the maximum marginal latency reduction to the KV mode.
Conversely, the backward search initializes with the ``All-KV" state ($\boldsymbol{x} = \mathbf{1}$, communication-intensive) and iteratively reverts AAs to the NL mode.
This bidirectional approach explores the solution space from opposite directions, significantly enhancing the probability of escaping local optima.

\emph{2) Inner Kernel (Optimal Bandwidth Allocation):} 
For every candidate mode vector $\boldsymbol{x}$ generated in the outer loop, its corresponding optimal bandwidth allocation $\boldsymbol{\rho}^{\star}$ and system latency can be derived via Bisection Search.
This ensures that every discrete transition in the outer loop is evaluated under its best possible continuous resource allocation, providing a rigorous metric to guide the greedy selection.

\emph{3) Decision and Termination:} 
At each iteration, the algorithm selects the mode transition that yields the steepest drop in marginal latency. 
The iterative process terminates when no further latency reduction can be achieved or the predefined state space is fully traversed. 
Finally, the best configuration is selected from the trajectories of both search directions, yielding a high-quality near-optimal joint solution.

{
\textbf{Complexity Analysis:}
The computational complexity of the proposed JMSRA algorithm is evaluated by analyzing its nested architecture.
For the inner continuous bandwidth allocation, the bisection search operates over a target latency. 
Let $\Delta T$ denote the initial latency search range and $\epsilon$ be the error tolerance. 
The number of bisection iterations is strictly bounded by $\log_2(\Delta T / \epsilon)$. 
Within each iteration, evaluating the feasibility of the trial latency requires summing the required bandwidths across all $I$ agents, incurring an operation cost of $\mathcal{O}(I)$. 
Thus, the complexity of a single inner loop execution is $\mathcal{O}\left(I \log_2(\frac{\Delta T}{\epsilon})\right)$.
For the outer discrete mode selection, the bidirectional greedy search requires evaluating multiple candidate mode vectors. 
In the worst-case scenario, the forward search evaluates $I + (I-1) + \dots + 1 = \frac{I(I+1)}{2}$ candidates to fully traverse the state space, yielding a complexity of $\mathcal{O}(I^2)$. 
The backward search introduces an identical worst-case overhead.
By synthesizing the nested layers, the overall worst-case computational complexity of the proposed JMSRA algorithm is given by $\mathcal{O}\left( I^3 \log_2\left(\frac{\Delta T}{\epsilon}\right) \right).$
Moreover, it is worth emphasizing that the practical complexity is lower than this theoretical worst-case bound. 
In practical executions, the bidirectional search avoids exhausting the maximum search depth of $I$ state transitions in both directions. 
Since either search direction halts independently the moment it yields no further latency reduction, a massive number of redundant candidate evaluations are skipped. 
Consequently, the actual number of transition steps executed is frequently less than the theoretical limit, as verified in \cref{fig:performance_case_study} (a) and (b).
Compared to the exhaustive search benchmark, whose complexity scales exponentially as $\mathcal{O}\left( 2^I \cdot I \log_2(\frac{\Delta T}{\epsilon}) \right)$, the proposed JMSRA algorithm exhibits a low polynomial complexity, enhancing its practical applicability in large-scale MASs.
}

\subsection{Numerical Results}

{
To validate the effectiveness of the proposed scheme, we povide the following numerical results.
\begin{itemize}
    \item JMSRA for AAs-to-EA Link (\cref{fig:performance_case_study,fig:performance_case_study2}): Aiming to validate the validity of the proposed JMSRA algorithm, we illustrate the mode allocation topologies assigned to heterogeneous AAs under varying system parameters, and evaluate the marginal latency compared with baseline schemes.
    \item Mode Switching for Multi-Round Dialogues (\cref{fig:performance_case_study3}): Subsequently, we extend the simulation to a multi-turn conversational scenario that better reflects the dynamics of real-world embodied environments, encompassing the complete interaction loop (i.e., both EA-to-AAs and AAs-to-EA phases).
\end{itemize}
}

\subsubsection{{Validation of JMSRA}}

\begin{table}[!h]
    \footnotesize
    \caption{Simulation Parameters for JMSRA Validation}
    \label{tab:sim_params1}
    \tabcolsep 0pt 

    \begin{tabular*}{\textwidth}{@{\hspace{1.5cm}} l @{\extracolsep{\fill}} l @{\extracolsep{\fill}} l @{\hspace{1.5cm}}}
        \toprule
        \textbf{Parameter Description} & \textbf{Symbol} & \textbf{Value} \\
        \midrule

        \multicolumn{3}{c}{\emph{Agent Capabilities and Wireless Environment Settings}} \\ \midrule
        Bits per token                         & $b$                                            & 16 \\
        KV cache compression ratio             & $\gamma_i, i\in\{1, \cdots, I\}$               & 2 \\
        Transmission bandwidth                 & $B$                                            & 2 GHz \\
        Noise power spectral density           & $N_0$                                          & -140 dBm/Hz \\
        Distance between EA and AA             & $d_i, i\in\{1, \cdots, I\}$                    & $\sim \mathcal{U}[5, 10]$ m \\
        Path loss                              & $PL(d_i)$                                      & $30 + 35 \log_{10}(d_i)$ dB \\
        Small-scale fading                     & $h_i$                                          & Rayleigh (Unit variance) \\
        Maximum transmit power                 & $P_i, i\in\{1, \cdots, I\}$                    & $\sim \mathcal{U}[10, 23]$ dBm \\
        \midrule

        \multicolumn{3}{c}{\emph{Single-Round Task Specific Parameters}} \\ \midrule
        Generated tokens length                & $\alpha_i, i\in\{1, \cdots, I\}$               & $\sim 1024 \times (0.8 + 0.4 \times \mathcal{U}[0, 1])$ \\
        Unshared context length of KV cache    & $\xi_i, i\in\{1, \cdots, I\}$                  & $\sim \mathcal{U}[2.8, 3.2] \times \alpha_i$ \\
        Historical context length              & $\theta_{0}$                                   & $1024 \times 5$ \\
        \bottomrule
    \end{tabular*}
\end{table}

We consider an embodied agent network comprising one EA and $I$ AAs. To comprehensively capture the heterogeneity and dynamics of realistic environments, the simulation parameters are randomized using a fixed random seed to ensure reproducibility. 
The specific numerical settings are summarized in \cref{tab:sim_params1}.

\emph{1) Network Topology and Channel Model:} 
The AAs are spatially distributed in an annular region centered at the EA. The wireless channel accounts for both large-scale path loss and small-scale Rayleigh fading to simulate realistic stochastic signal degradation.

\emph{2) Heterogeneous Agent Capabilities:}
To reflect the diversity of hardware capabilities in practical robotic scenarios, we introduce uniform heterogeneity in both the transmit power and computing capacity among the AAs. 
Meanwhile, the computing capability of the EA, i.e., $C_0$, serves as a tunable parameter in our evaluation to investigate the system behavior under different computational bottlenecks.

\emph{3) LLM and Task Settings:}
The specific LLM parameters are the same as those in \cref{tab:sim_params}.
Crucially, we assume a homogeneous deployment where all embodied agents utilize the identical LLM architecture and weights. This assumption is necessary because the direct transmission of KV cache is natively supported only between identical models. In contrast, collaborative inference among heterogeneous LLMs would require additional feature alignment or projection mechanisms to ensure semantic understanding, which introduces complexity beyond the scope of this latency-centric study. Therefore, we reserve the investigation of heterogeneous model interoperability and the impact of KV cache quantization on semantic accuracy for future work.
Regarding the inference tasks, the output token length and the corresponding unshared context length are modeled as random variables to simulate diverse inference scenarios in real-world queries. 
It is worth noting that practical agents are subject to finite memory constraints (e.g., KV cache limits), where excessive context lengths may lead to task failures. However, given that the primary focus of this work is the trade-off between computation and communication latencies, we assume that the assigned tasks fall within the feasible memory capacity of the agents. The rigorous analysis of memory-constrained collaborative inference is reserved for future work.

\begin{figure}[!t]
    \centering

    \subfloat[]{
        \label{fig:algo_lowC0} 

        \includegraphics[height=6.06cm, keepaspectratio]{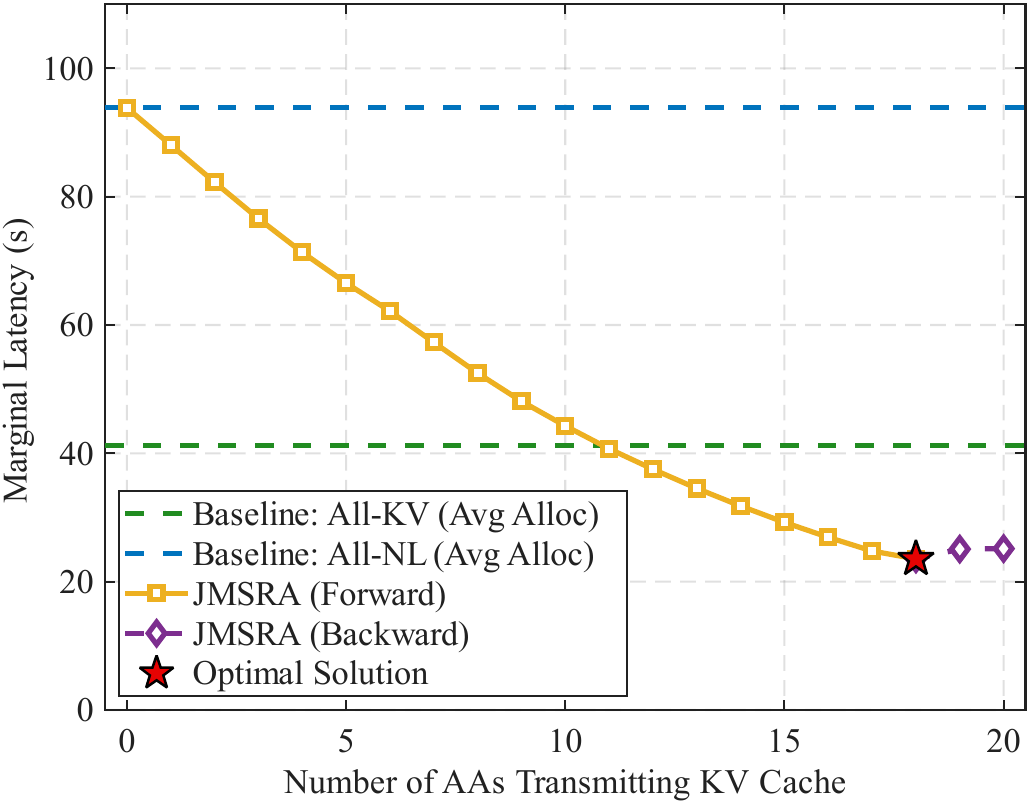}
    }
    \hfil
    \subfloat[]{
        \label{fig:algo_highC0}
        \includegraphics[height=6.06cm, keepaspectratio]{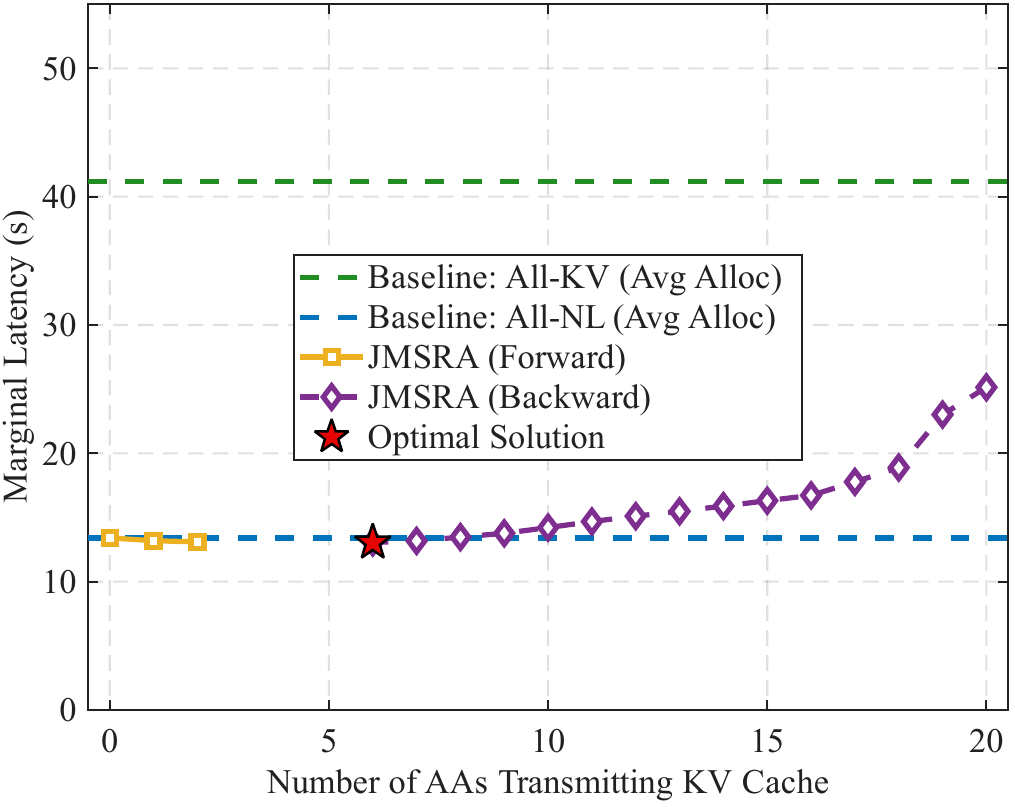}
    }
    \\
    \subfloat[]{
        \label{fig:pos_lowc0}
        \includegraphics[height=6.4cm, keepaspectratio]{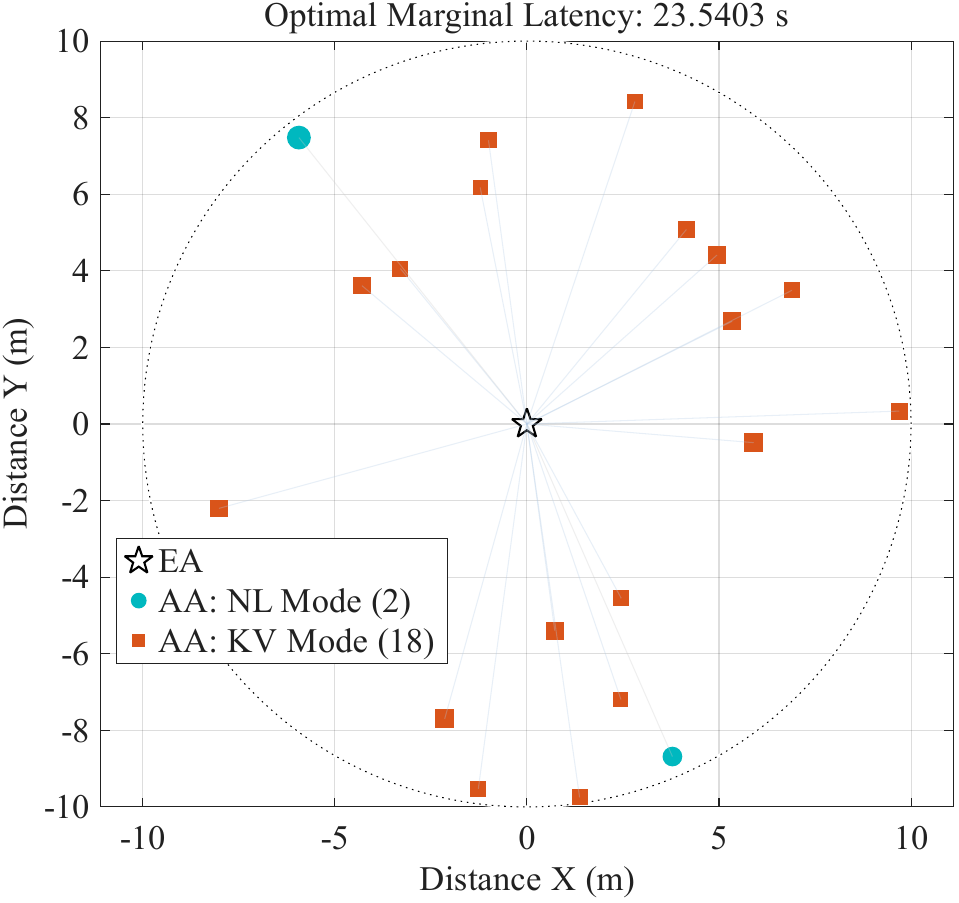}
    }
    \hspace{0.06\textwidth} 
    \subfloat[]{
        \label{fig:pos_highC0}
        \includegraphics[height=6.4cm, keepaspectratio]{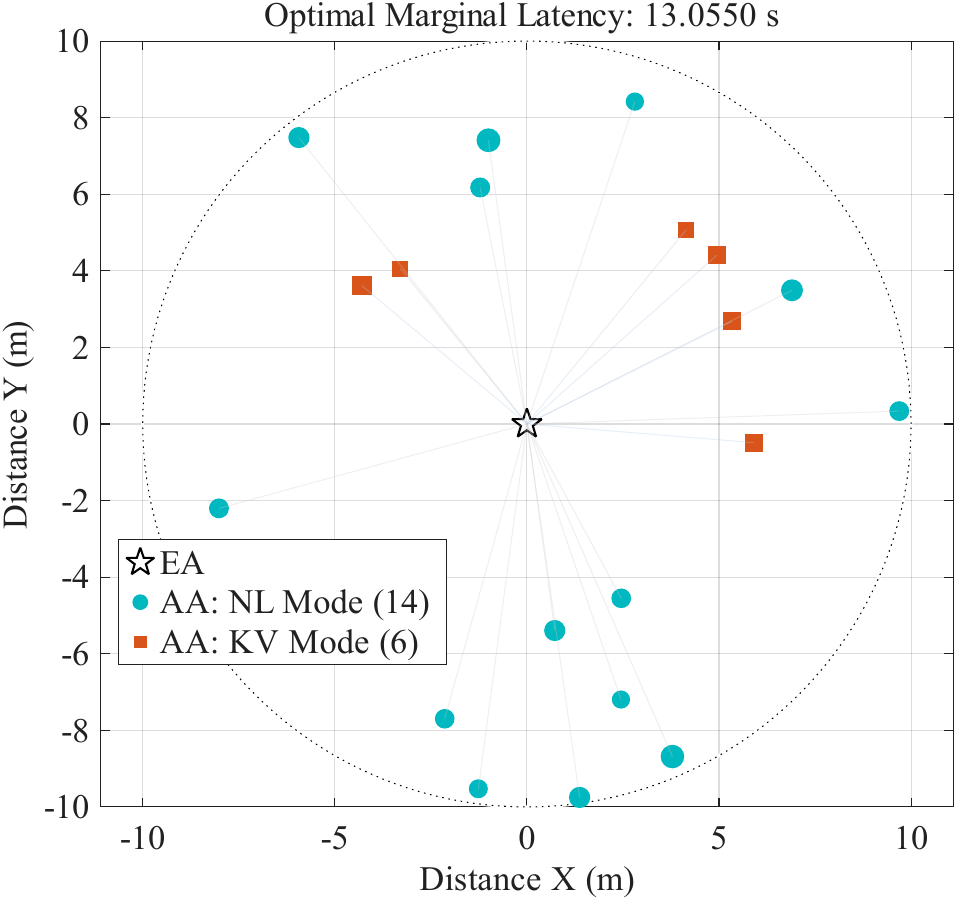}
    }
    
    \caption{Performance of the proposed algorithm, where (a), (c) $C_0$ = 5 TFLOPS, and (b), (d) $C_0$ = 35 TFLOPS. Here, we set $I=20$.}
    \label{fig:performance_case_study} 
\end{figure}

\cref{fig:performance_case_study} (a) and (c) respectively illustrate the latency optimization footprint and the resulting mode selection when EA's computing capacity is limited.
Observed from \cref{fig:performance_case_study} (a), the proposed algorithm outperforms both baselines (with uniform bandwidth allocation), demonstrating its effectiveness in balancing computation and communication loads.
The ``All-NL" baseline suffers from an excessively high latency of over 90 seconds while the ``All-KV" baseline yields a much lower latency of around 41 seconds.
This reveals that the EA's limited computing power creates a severe bottleneck during the prefill phase when handling token inputs from all AAs. 
As the JMSRA algorithm (forward path) progressively switches agents to the KV mode, the marginal latency decreases monotonically. 
Besides, in the backward path, the algorithm starts from the ``All-KV" state and iteratively reverts agents to NL mode. 
One can see that the forward and backward search paths converge to the same optimal latency point, which is 23.54 s with 18 AAs in KV mode, achieving about $75\%$ reduction compared to the ``All-NL" baseline.
In the meantime, \cref{fig:performance_case_study} (c) shows the mode selection result of \cref{fig:performance_case_study} (a). 
It can be seen that the network is dominated by agents utilizing KV cache transmission mode. 
Furthermore, it can be observed that most AAs closer to EA are assigned to KV mode. This is because the large-scale channel fading between these AAs and EA is relatively small, thus providing better channel capacity to support the transmission of KV cache.

\cref{fig:performance_case_study} (b) and (d) depict the system performance and mode selection when the EA is equipped with sufficient computing capacity.
In this case, the ``All-NL" baseline achieves a low latency of approximately 13.5 seconds, while the ``All-KV" baseline suffers from a high latency of around 41 seconds due to the heavy communication overhead of transmitting KV cache. 
The proposed JMSRA algorithm identifies a operating point at 13.05 seconds, slightly outperforming the ``All-NL" baseline.
This indicates that when the EA is powerful enough to handle tokens coming from the AAs, the system favors the NL mode to minimize communication latency.
It is noteworthy that the efficacy of bandwidth optimization is less pronounced in this regime. 
With the majority of AAs operating in the NL mode, the transmission load is significantly reduced relative to the KV mode in a sufficient bandwidth environment.
In addition, one can observe from \cref{fig:performance_case_study} (b) that the forward and backward search paths stop at different local optima, with the backward path achieving a slightly better latency.
This shows the advantage of bidirectional search in avoiding poor local minima.
Furthermore, it can be seen that, compared to \cref{fig:performance_case_study} (c), those AAs in the KV mode in \cref{fig:performance_case_study} (d) are closer to EA.
This further confirms that the proposed algorithm can effectively leverage the channel conditions to guide the mode selection, thus optimizing the overall system performance.
{
Such a capability can play a pivotal role in practical scenarios.
Consider, for instance, an unmanned swarm for disaster relief, where a mobile command vehicle (EA) coordinates reconnaissance UAVs (AAs).
For UAVs with favorable line-of-sight channels, JMSRA dynamically selects the KV mode to transfer rich semantic insights while maintaining low E2E latency. 
Conversely, for UAVs severely shadowed by collapsed structures, it adaptively switches to the highly compressed and more robust NL mode. 
}

\begin{figure}[!t]
    \centering

    \subfloat[]{
        \label{fig:perf_B} 
        \includegraphics[height=6.2cm, keepaspectratio]{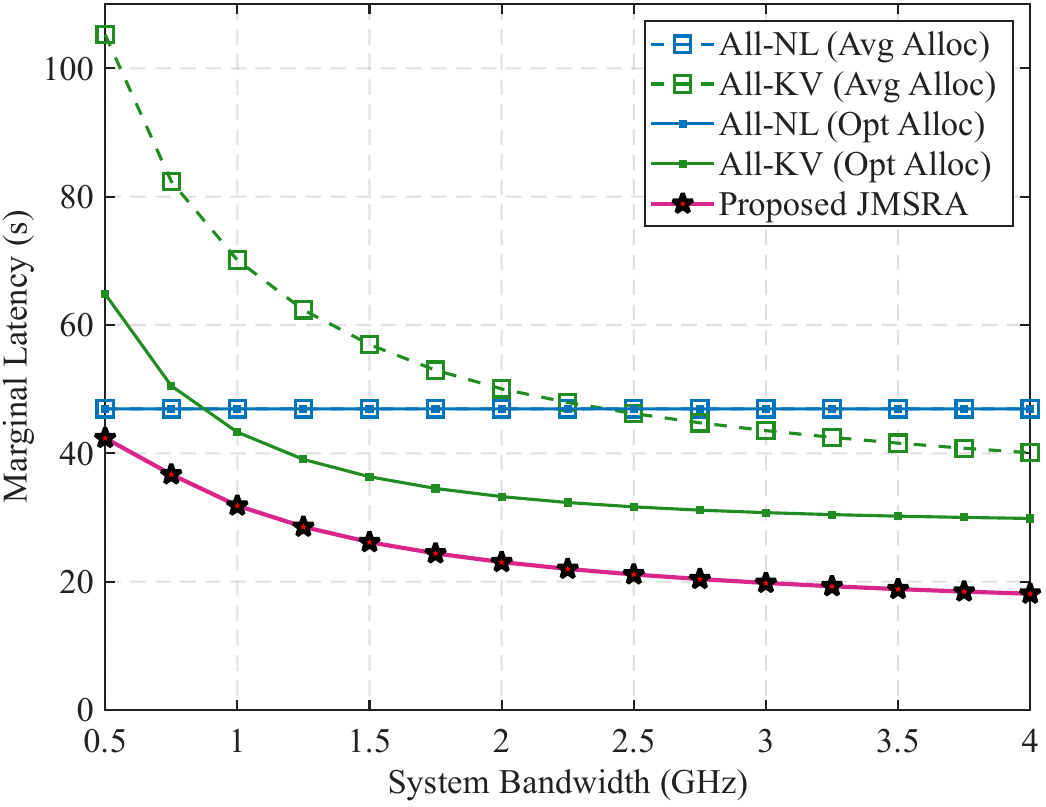}
    }
    \hfil 
    \subfloat[]{
        \label{fig:perf_I}
        \includegraphics[height=6.2cm, keepaspectratio]{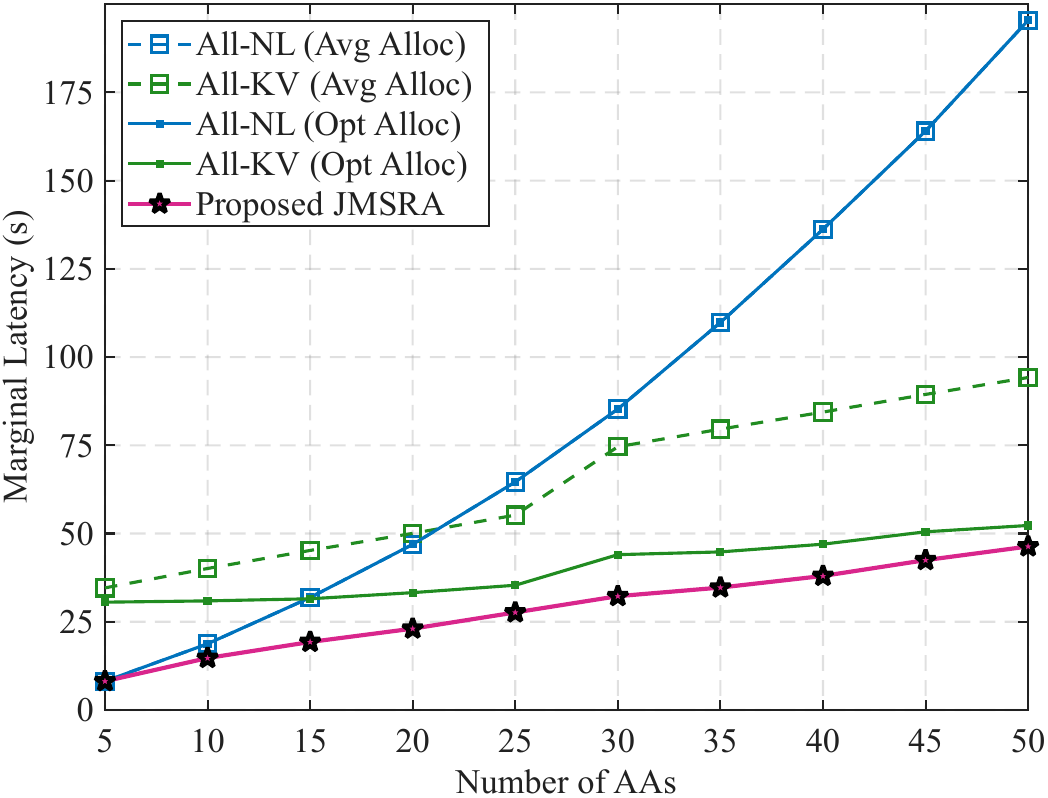}
    }
    \caption{Performance of the proposed scheme versus (a) total bandwidth ($I = 20$) and (b) number of AA ($B = 2$ GHz). Furthermore, both figures set $C_0 = 10$ TFLOPS.}
    \label{fig:performance_case_study2} 
\end{figure}

It is observed from \cref{fig:performance_case_study2} (a) that the proposed JMSRA scheme achieves the lowest latency across the entire bandwidth conditions. 
As the system bandwidth $B$ increases, the latency decreases in all regimes except for those with low transmission load.
For the ``All-NL'' baselines, both the average and optimized allocation schemes exhibit identical, flat performance curves. 
This is because the ``All-NL'' mode is computation-bound but not communication-bound, unless the transmitted token volume is large enough.
In contrast, for the ``All-KV'' baselines, a significant performance gap is observed between the average and optimized allocations. 
Since full KV mode is communication-bound, particularly in bandwidth-constrained regimes (e.g., $B < 2$ GHz), inefficient bandwidth allocation leads to severe congestion. 
The optimized allocation (``All-KV-Opt'') significantly mitigates this, validating the necessity of the resource allocation mechanism.
Most importantly, the proposed JMSRA demonstrates superior robustness. 
In both low and high bandwidth regime, the proposed JMSRA can successfully identify a transmission strategy to effectively optimize the system latency, avoiding the pitfalls of choosing a unified strategy.

One can see from \cref{fig:performance_case_study2} (b) that the proposed JMSRA scheme consistently maintains the lowest latency across all deployment scales. 
While the latency of JMSRA increases with the number of agents $I$, it exhibits a growing advantage over the ``All-NL'' baselines and will eventually parallel the ``All-KV-Opt'' scheme.
For the ``All-NL'' baselines, the latency increases sharply with $I$. 
This is attributed to the heavy computation load at the EA. 
Upon receiving feedback data from AAs, the EA must process these aggregated tasks, incurring a computation latency that scales quadratically with the input data volume.
In this compute-bound regime, bandwidth optimization provides negligible gains.
Conversely, for the ``All-KV'' baselines, the trend is dictated by communication overhead.
With average allocation, the latency rises relatively faster because the limited bandwidth is not optimally allocated, causing the system to be bottlenecked by the AA with the worst channel condition.
However, with bandwidth optimization, the curve is more flat than the average allocation curve. 
This is because the optimization algorithm dynamically balances the load, allocating more bandwidth to AAs with poor channels. 
It is worth noting that JMSRA performs almost the same as ``All-NL'' at low $I$ values (e.g., $I=5$), but its performance tends to be similar to ``All-KV-Opt'' at high $I$ values. 
This confirms again that the JMSRA can determine the transmission strategy based on the situation.

\subsubsection{{Mode Switching for Multi-Round Dialogues}}
\begin{table}[!h]
    \footnotesize
    \caption{Simulation Parameters for Multi-Round Dynamic Scenario}
    \label{tab:sim_params_multiturn}
    \tabcolsep 0pt 

    \begin{tabular*}{\textwidth}{@{\hspace{1cm}} l @{\extracolsep{\fill}} l @{\extracolsep{\fill}} l @{\hspace{1cm}}}
        \toprule
        \textbf{Parameter Description} & \textbf{Symbol} & \textbf{Value} \\
        \midrule
        \multicolumn{3}{c}{\emph{Agent Capabilities and Wireless Environment Settings}} \\ \midrule
        Available AAs                                                       & $\mathcal{I}_j$                                           & Changeable ($I_{\max} = 20$)\\
        EA computing capability                                             & $C_0$                                                     & 20 TFLOPS \\
        AA computing capability                                             & $C_i, i\in\{1, \dots, I_{\max}\}$                         & $\sim \mathcal{U}[5, 15]$ TFLOPS \\
        EA transmit power                                                   & $P_0$                                                     & 30 dBm \\
        AA transmit power                                                   & $P_i, i\in\{1, \dots, I_{\max}\}$                         & $\sim \mathcal{U}[10, 23]$ dBm \\
        Dynamic distance                                                    & $d_{i,j}, i\in\{1, \dots, I_{\max}\}$                     & $\sim \mathcal{U}[5, 10]$ m \\
        Path loss                                                           & $PL(d_{i,j})$                                             & $30 + 35 \log_{10}(d_{i,j})$ dB \\
        Small-scale fading                                                  & $h_{i,j}, i\in\{1, \dots, I_{\max}\}$                     & Rayleigh (Unit variance) \\
        \midrule
        \multicolumn{3}{c}{\emph{Multi-Round Task Specific Parameters}} \\ \midrule
        EA initial task prompt tokens                                       & $s_{0,1}$                                                 & 1024 \\
        EA generated tokens                                                 & $\alpha_{0,j}$                                            & 1024 \\
        AA sensory input tokens                                             & $s_{i,j}, i\in\mathcal{I}_j$                              & 1024 \\
        AA generated tokens                                                 & $\alpha_{i,j}, i\in\mathcal{I}_j$                         & 1024 \\
        EA context limit                                                    & $\theta_{0,\max}$                                         & $1024 \times I_{\max} \times 20$ \\
        EA sliding window size for unshared KV debt ($\xi_{0,j}$)           & $W_{0}$                                                   & $1024 \times I_{\max} \times 5$ \\
        AA context limit                                                    & $\theta_{i,\max}, i\in\{1, \dots, I_{\max}\}$             & $1024 \times 50$ \\
        AA sliding window size for unshared KV debt ($\xi_{i,j}$)           & $W_{i}, i\in\{1, \dots, I_{\max}\}$                       & $1024 \times 10$ \\
        \bottomrule
    \end{tabular*}
\end{table}

{While we have explored the transmission mode selection between the EA and the AAs in previous sections, the analysis above was limited to single-round dialogues. 
In the real world, multi-agent cooperative systems typically require multiple rounds of interaction. 
Therefore, this section will further evaluate the usability of the proposed scheme in dynamic multi-turn dialogue scenarios.}

\begin{figure}[!t]
    \centering

    \subfloat[]{
        \label{fig:multi_turn_heatmap} 
        \includegraphics[height=5.85cm, keepaspectratio]{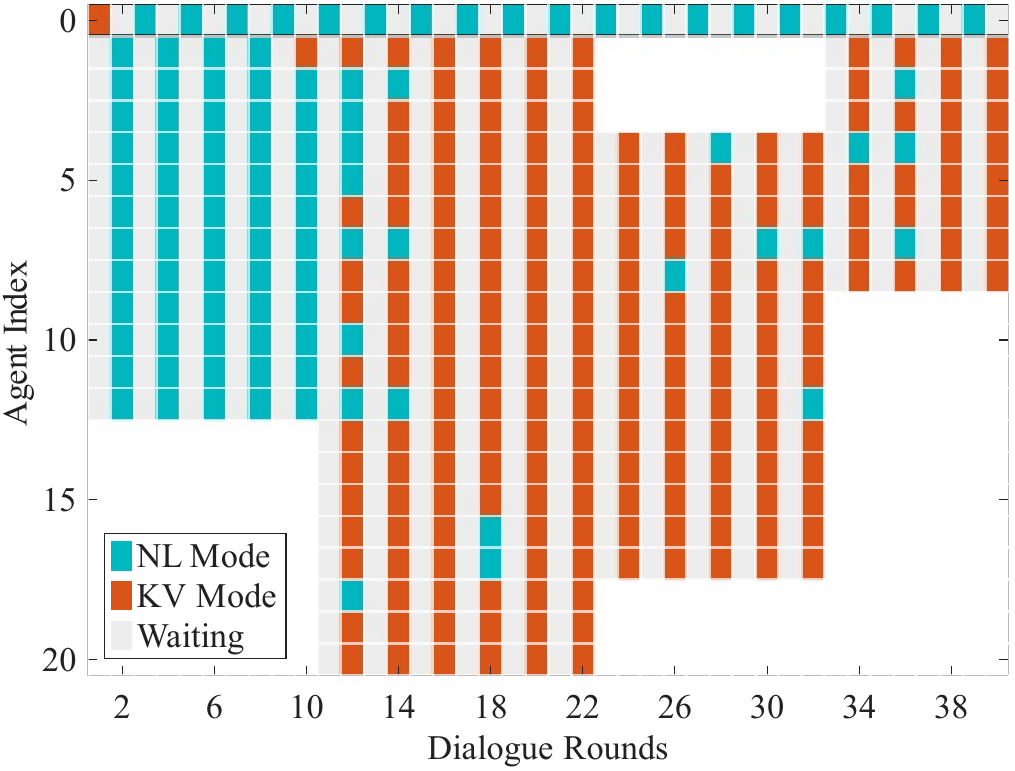}
    }
    \subfloat[]{
        \label{fig:multi_turn_EA_latency}
        \includegraphics[height=5.85cm, keepaspectratio]{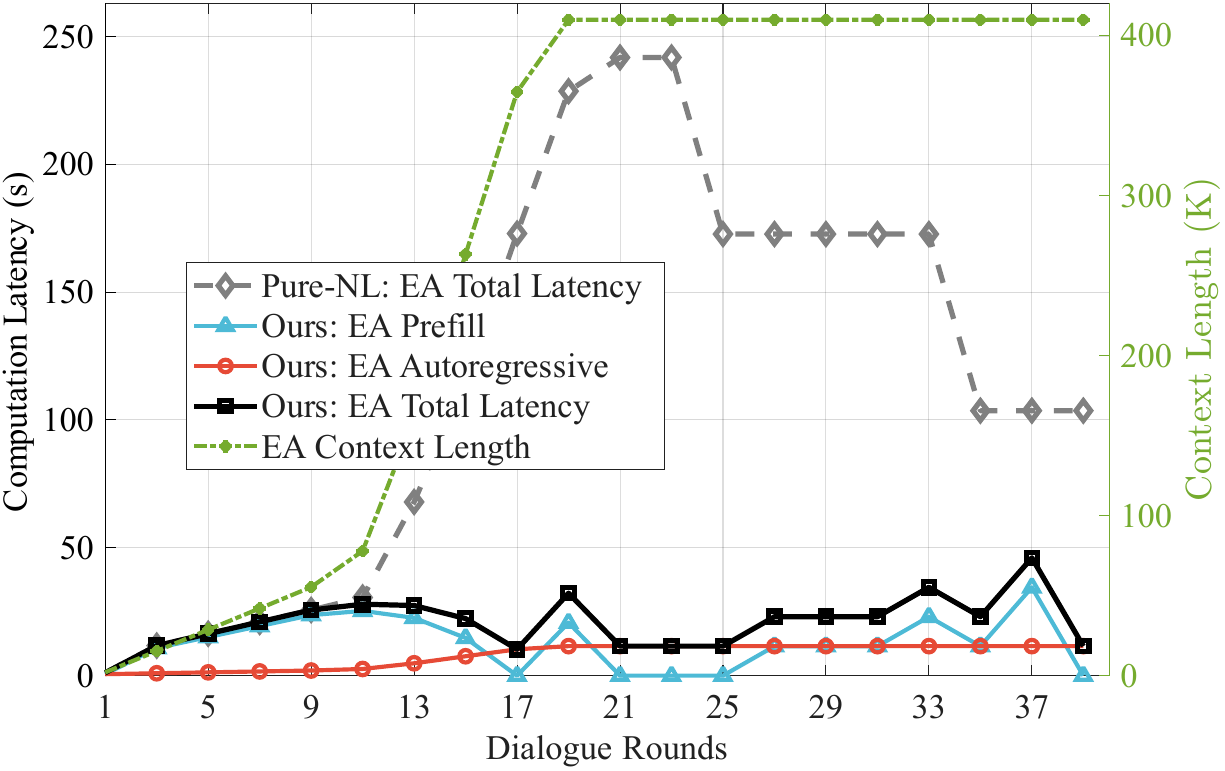}
    }

    \caption{Performance of the proposed scheme in a dynamic multi-round dialogue scenario. (a) Scheduling heatmap of the transmission modes across all active agents. (b) Internal-level breakdown of the EA inference performance in all rounds.}
    \label{fig:performance_case_study3} 
\end{figure}

{To simulate dynamic multi-turn scenarios, we set new simulation parameters, as shown in \cref{tab:sim_params_multiturn}. 
Other involved parameters are the same as stated before.
It is noteworthy that the number of active AAs changes dynamically in different dialogue rounds, reflecting a real cooperative environment.
For instance, some AAs may temporarily exit the current communication network to perform urgent tasks. 
Furthermore, the distance between the EA and each AA, as well as the wireless channel state, are updated independently in each dialogue round. 
Conversely, since the effectiveness of the proposed scheme in handling heterogeneous generation lengths has been verified in previous sections, we fix the single-turn input and output token lengths for all agents to 1024 in this subsection. 
This aims to eliminate the interference caused by drastic fluctuations in input or output length during a single round, thereby allowing for a clearer and smoother observation of the cumulative performance trend in multi-round interactions (as shown in \cref{fig:performance_case_study3} (b)).
}

{
Particularly, for each agent, we define a context window limit, as well as a sliding window to limit unshared KV cache debt. 
Since unshared KV cache accumulates continuously in consecutive dialogues, without effective limits, large KV debts will generate extreme transmission latency, ultimately causing the mode-switching mechanism to fail. 
Therefore, during the decision phase of transmission strategy, the amount of unshared KV cache that an agent can transmit is strictly constrained within this sliding window. 
Notably, the values of $\xi_{0,j}$ and $\xi_{i,j}$ are no longer random variables, but are calculated according to the actual dialogue process.
Finally, the debt clearing mechanism in the system must be emphasized: once an agent is scheduled to successfully transmit its KV cache in a specific round, its locally accumulated KV debt will be immediately cleared to zero, and recalculation will begin in subsequent dialogues.
}

{As shown in the system scheduling heatmap of \cref{fig:performance_case_study3} (a), the proposed scheme demonstrates superior adaptive trade-offs between computation and communication in the face of highly dynamic physical environments.
In the initial stage of multi-round interactions (e.g., the first 10 rounds), due to the relatively small historical context $\theta_0$ accumulated by the EA, the computational capability in the prefill stage is relatively abundant. 
The system tends to allow active AAs to adopt the NL mode, thereby greatly saving valuable wireless communication bandwidth. 
However, as the dialogue rounds deepen, continuing to allow AAs to adopt NL mode will cause a rapid increase in the prefill attention computation time at the EA. 
Therefore, the JMSRA algorithm decisively switches most active AAs to KV Mode in subsequent rounds. 
This avoids the huge prefill computation latency of the EA by sacrificing some bandwidth. 
Furthermore, as can be seen from \cref{fig:performance_case_study3} (a), the EA only selects KV mode in the first round, while adopting NL mode in all subsequent rounds. 
This is because the EA aggregates data from all AAs, resulting in a large KV cache debt within it, leading to a sharp increase in transmission latency, thus favoring NL mode. 
It is worth mentioning that although most AAs adopt KV mode in deep round dialogue, some AAs adopt NL mode, which further demonstrates the superiority of the proposed JMSRA algorithm.}

{\cref{fig:performance_case_study3} (b) breaks down the inference process within the EA from an internal perspective. 
\cref{fig:performance_case_study3} (b) intuitively demonstrates the advantages of the proposed scheme in suppressing the ``computational disaster" of LLMs in multi-agent collaborative scenarios. 
It can be clearly observed that as the dialogue progresses, the historical context $\theta_0$ of the EA continuously accumulates and is eventually limited to about 400K tokens. 
With this huge context base, if the AAs adopt the NL scheme in each round, the computation time of the EA's prefill phase will increase significantly, causing the total computation latency of the EA to soar to an unbearable 241 seconds in the 21st round.
Conversely, although the autoregressive latency continues to increase with the depth of the dialogue rounds, thanks to the JMSRA algorithm's switching of some AAs to KV mode and reasonable bandwidth allocation at key time nodes, the EA prefill computation latency of this scheme is consistently suppressed to an extremely low level throughout the all rounds of interaction, resulting in a total inference latency far lower than pure NL. 
This strongly demonstrates the superior performance of the proposed scheme in maintaining the long-term viability of LLM-driven multi-agent collaboration.
}

\section{Conclusion and Future Work}
This paper investigated communication media selection and bandwidth allocation strategies for wireless-enabled multi-agent collaboration.
We elevated traditional communication media to the semantic or cognitive level, focusing on two distinct transmission modes: token-based and KV-cache-based modes.
By mathematically modeling the underlying computation and communication processes, we established the E2E latency expressions for both token-based and KV-cache-based transmission modes.
Simulation results demonstrated that neither mode consistently outperforms the other, as their efficiency heavily depends on the specific system parameters.
To capture the potential minimum system latency, we proposed mode selection strategies from the perspectives of both individual agents and multi-agent resource competition.
For the single-agent case, we derived a closed-form expression for the optimal mode switching point.
For the multi-agent case, we formulated a joint optimization problem for mode selection and bandwidth allocation, and developed an efficient JMSRA algorithm to solve it.
The JMSRA algorithm utilized a low-complexity bi-directional greedy approach that avoids exhaustive search by initiating from both all-NL and all-KV configurations. 
Numerical results confirmed that the proposed scheme consistently outperforms static NL-only and KV-cache-only baselines across diverse environments, while maintaining excellent long-term stability in dynamic multi-round dialogues.
These findings provided critical insights from a wireless communication perspective for deploying collaborative embodied agents in industrial environments and beyond.

{
While this paper evaluated media switching through the lens of latency reduction, its implications for the task success rate also provide an interesting perspective.
Recent natural language processing studies indicate that in multi-agent collaboration scenarios, token-based interactions suffer from semantic compression and error accumulation over multiple dialogue turns. 
Exchanging KV caches, however, allows agents to preserve high-dimensional cognitive fidelity, thereby capable of achieving higher task success rates than token-based approaches.
Nevertheless, when transitioning from the internal computing environment to the physical wireless world, this characteristic is reversed. 
The stochastic nature of wireless channels inevitably introduces signal perturbations. 
For discrete tokens, despite the impact of fluctuating channel quality, they enter the LLM's prefill stage, endowing them with a certain degree of contextual self-correction capability to resist physical-layer bit errors. 
Conversely, because KV caches are directly embedded into the memories of receiving agents and they are exceptionally vulnerable to channel perturbations, even minor signal disturbances can trigger severe semantic hallucinations during the autoregressive generation process.
Such an intriguing trade-off making extending the proposed mode selection strategy to incorporate task-level reliability a fascinating future direction. 
Furthermore, the quantization of the KV cache serves as another critical parameter closely coupled with the media switching decision. 
When channel conditions support KV-cache transmission, the system can dynamically enhance quantization precision or even employ complex non-uniform quantization to compensate for channel fluctuations. 
Conversely, aggressive quantization curtails latency but inevitably leads to a decline in the task success rate. 
Therefore, future work can integrate KV-cache quantization into the proposed framework to strike an optimal balance between semantic fidelity and physical-layer performance. 
Additionally, to improve generality, extending the JMSRA framework to heterogeneous agent scenarios where different agents deploy distinct LLM architectures represents another vital direction. 
This will necessitate investigating cross-model KV-cache feature alignment and token vocabulary mapping to achieve seamless semantic interoperability across diverse models.
Moreover, such architectural heterogeneity introduces geometric discrepancies in KV cache structures and leads to asymmetric latency models across diverse agents. 
Future extensions must therefore accommodate these structural and mathematical non-uniformities to dynamically optimize resource allocation under non-homogeneous nodes.
}








\bibliographystyle{scis}
\bibliography{ref}






\end{document}